\documentclass[sigconf]{acmart}

\AtBeginDocument{%
  \providecommand\BibTeX{{%
    \normalfont B\kern-0.5em{\scshape i\kern-0.25em b}\kern-0.8em\TeX}}}

\usepackage{array}
\usepackage[linesnumbered,ruled,vlined]{algorithm2e}
\usepackage{subfigure}
\usepackage{diagbox}
\usepackage{multirow}
\usepackage{booktabs}
\usepackage{color}

\newcommand\our{\text{FDLoRA}}

\setcopyright{acmlicensed}
\copyrightyear{2018}
\acmYear{2018}
\acmDOI{XXXXXXX.XXXXXXX}

\acmConference[Conference acronym 'XX]{Make sure to enter the correct
  conference title from your rights confirmation emai}{June 03--05,
  2018}{Woodstock, NY}
\acmISBN{978-1-4503-XXXX-X/18/06}

\begin{document}

\title{FDLoRA: Personalized Federated Learning of Large Language Model via Dual LoRA Tuning}

\author{
Yao Lu$^1$$^\dagger$,
Jiaxing Qi$^1$$^\dagger$,
Zhongzhi Luan$^1$,
Shaohan Huang$^1$,
Carol Fung$^2$,
Hailong Yang$^1$,
Depei Qian$^1$
}
\thanks{$\dagger$ These authors contributed equally to this work.}
\thanks{Corresponding authors: \texttt{\{jiaxingqi, luan.zhongzhi\}@buaa.edu.cn}}

\affiliation{
  \institution{Sino-German Joint Software Institute, Beihang University, Beijing, China$^1$}
  \country{}
}
\affiliation{
  \institution{Concordia Institute for Information Systems Engineering, Concordia University, Montreal, Canada$^2$}
  \country{}
}

\begin{abstract}
Large language models (LLMs) have emerged as important components across various fields, yet their training requires substantial computation resources and abundant labeled data. It poses a challenge to robustly training LLMs for individual users (clients). To tackle this challenge,  the intuitive idea is to introduce federated learning (FL), which can collaboratively train models on distributed private data. However, existing methods suffer from the challenges of data heterogeneity, system heterogeneity, and model size, resulting in suboptimal performance and high costs. In this work, we proposed a variant of personalized federated learning (PFL) framework, namely \our{}, which allows the client to be a single device or a cluster and adopts low-rank adaptation (LoRA) tuning. \our{} sets dual LoRA modules on each client to capture personalized and global knowledge, respectively, and only the global LoRA module uploads parameters to the central server to aggregate cross-client knowledge. Finally, an adaptive fusion approach is employed to combine the   of the dual LoRAs. This enables \our{} to make effective use of private data distributed across different clients, thereby improving performance on the client without incurring high communication and computing costs. We conducted extensive experiments in two practice scenarios. The results demonstrate that \our{} outperforms six baselines in terms of performance, stability, robustness, computation cost, and communication cost.
\end{abstract}



\keywords{Personalized Federated Learning, LLM, LoRA Tuning, Low Communication}


\received{20 February 2007}
\received[revised]{12 March 2009}
\received[accepted]{5 June 2009}

\maketitle

\section{Introduction}
Recently, large language models (LLM), such as ChatGPT \cite{wu2023brief} and LLaMA \cite{touvron2023llama}, have gained significant attention across various domains. As LLMs become more prevalent, tailoring them to specific tasks becomes increasingly necessary. For example, LLM can be customized to analyze local medical data for different institutions in the medical scenario \cite{sheller2020federated}. This customization often relies on supervised fine-tuning (SFT), a process where LLMs are adapted to domain-specific data through various techniques \cite{kim2023efficient}. The current primary SFT technologies can be categorized into three groups \cite{wu2024fedlora}: fine-tuning, parameter-efficient fine-tuning (PEFT), and prompt-tuning.
LoRA has attracted attention for its effectiveness in PEFT. By optimizing the rank-decomposition matrix, LoRA can train low-rank parameter matrices in LLMs while freezing the original parameters (see Section \ref{sec:preliminaries}). LoRA not only enhances the performance of LLMs on domain-specific tasks but also mitigates the need for extensive labeled data \cite{hu2021lora}. This is particularly valuable and efficient in scenarios where such data and computational resources are scarce. 
\par
Let us consider the following scenario: Internet service providers (ISPs) collect a large amount of logs to record network activity, system events, and potential threats. However, the cost of manually analyzing these logs is huge. Recently, there has been a lot of work on using LLM to analyze private logs and determine whether an anomaly occurs in the system \cite{qi2023loggpt, xu2024unilog}, which requires manual labeling of a large amount of logs. Although the goal of ISPs is to use LLM to analyze private logs to improve system reliability, the limited number of labeled logs is an obstacle to the effective use of the LLM capability. In addition, because log data is very valuable and private, ISPs will not share the private logs with each other.  Therefore, the problem arises: \textit{How to take advantage of LLM in similar practical application scenarios?} 
\par
The integration of FL with LLMs offers a promising opportunity for collaborative learning across clients \cite{mcmahan2017communication, zhang2024towards}. However, the inherent challenge of data heterogeneity, where the distribution of data across clients is non-identically and independently distributed (\textit{non-IID}), poses significant obstacles \cite{ma2022layer, chen2023fedsoup}. This often leads to suboptimal performance of the global model due to convergence to the worst performing client, referred to as the "\textit{bucket effect}" \cite{bai2024federated}. To address this challenge, recent efforts have combined personalized federated learning (PFL) with techniques such as LoRA tuning to train personalized local models for each client and mitigate the traditional "\textit{bucket effect}" \cite{yi2023fedlora, wu2024fedlora}. The PFL aims to tailor the training process to individual clients, taking into account their unique data distributions and learning objectives, which allows the model to capture  not only cross-client knowledge (i.e., global/shared knowledge) but also client-specific knowledge (i.e., local/personalized knowledge). By personalizing the training process, PFL seeks to mitigate the negative effects of data heterogeneity and improve the performance of each client \cite{huang2021personalized}. Nonetheless, the computational and communication overhead associated with handling LLM parameters, which rapidly increasing from millions to billions, remains a significant challenge for these methods.

\par
To tackle these challenges, we take inspiration from literature on previous works \cite{huang2023lorahub, douillard2023diloco}, to integrate the LLM customization process into a variant of PFL framework. By doing so, we can gain insight from logs from different ISPs while maintaining privacy and improving performance. Each ISP acts as a client and contributes labeled data to LLM training without directly sharing sensitive information, thereby improving the stability of their respective systems by effectively leveraging distributed data sources.
\par
In this paper, we proposed a fusion-enhanced and low communication framework for LLM customization by integrating the LoRA tuning and variant PFL, a.k.a \textbf{\our{}}. In variant PFL, there are $N$ clients, each client uses private data to fine-tune LLM locally, and each client updates a replica of LLM. Instead of performing only one update, each client performs several updates locally. Then, exchanging parameters at each $H$-step to bring their parameters back into sync. In particular, the framework initializes two LoRA modules for LLM in each client, the personalized LoRA module and the global LoRA module, which capture personalized and global knowledge, respectively. The personalized LoRA module will not share parameters with other clients, thus ensuring that LLM adapts to local data. The global LoRA module uploads parameters to the server for aggregation operations, thereby integrating the knowledge of different clients. Furthermore, in order to preserve the basic knowledge learned by the LLM from the large corpus, we freeze the parameters of the LLM. Finally, we utilized a gradient-free optimization technique to adaptively fuse the parameters of both LoRA modules, enhancing the performance of the client's local task. Our primary contribution can be summarized as follows:

\begin{itemize}
    \item We proposed \our{}, a novel fusion-enhanced and low communication framework. To the best of our knowledge, this is the first work that integrates dual LoRA and variant PFL for LLM customization.
    \item We utilized both the LoRA module and gradient-free adaptive fusion approach, which effectively captured personalized and global knowledge and improved the performance of each client. 
    \item We conducted experiments in log analysis and medical diagnosis, demonstrating superior performance and stability. Our method can effectively reduce computation and communication costs.
\end{itemize}

\begin{table}[!tb]
\centering
\caption{Notations of problem settings.}
\label{tab:notation}
\begin{tabular}{ll} 
\toprule
\textbf{Notation} & \textbf{Explanation}  \\ 
\midrule
  $\mathcal{D}$       &    The collective dataset across all clients.     \\
  $(x, y)$            &    The data $x$ with label $y$.                   \\
  $N$                 &    The number of clients.                         \\
  $f(\cdot; \theta)$  &    The model with parameter $\theta$.              \\
  $\theta_{p}$        &    The trainable parameters of personalized part.    \\
  $\theta_{s}$        &    The trainable parameters of the shared (global) part.  \\
  $\mathcal{L}(\hat{y}, y)$        &    The loss function between $\hat{y}$ and $y$.  \\
  $\mathcal{G} (\cdot; \mathbf{w})$   &    The learnable parameters of fusion function $\mathcal{G}$.  \\
  
\bottomrule
\end{tabular}
\end{table}

\par
The rest of this paper is organized as follows. In Section \ref{sec:preliminaries}, we introduce the background and techniques related to our work. In Section \ref{sec:overview}, we put forward our framework and describe the design in detail. In Section \ref{sec:evaluation}, we demonstrate extensive experiments to evaluate the effectiveness of the proposed method in comparison with other methods. In Section \ref{sec:related}, we review  the related works of PFL and PEFT. Finally, we conclude our work in Section \ref{sec:conclusion}.

\section{Preliminaries} \label{sec:preliminaries}
In this section, we give a brief introduction to the problem definition of personalized federated learning and parameter-efficient fine tuning with low-rank adaptation. The meanings of some notations used in subsequent sections are declared in Table \ref{tab:notation}.

\begin{figure}[!tb]
    \centering
    \includegraphics[scale=0.45]{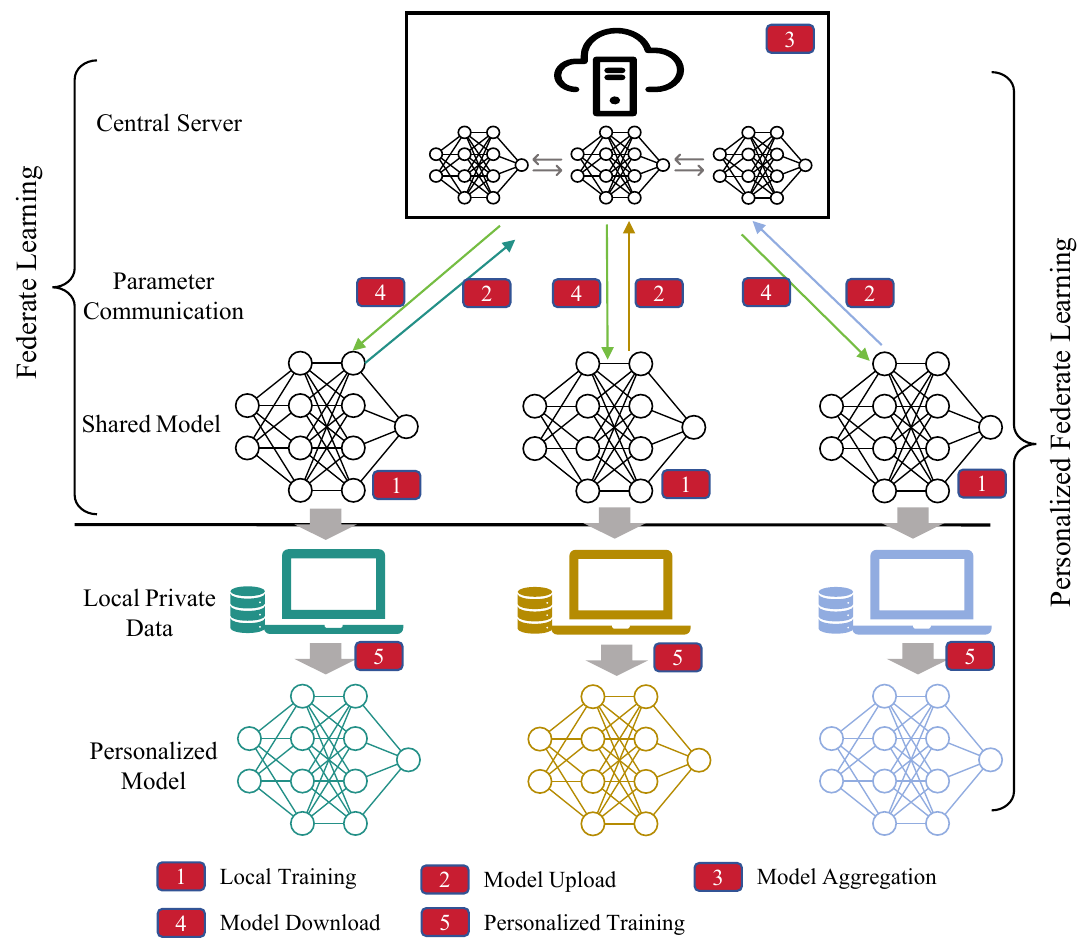}
    \Description[Short description]{Long description of the image.}
    \caption{The illustration of both federated learning and personalized federated learning. Personalized federated learning aims to find a personalized model for each client.}
    \label{fig:intro}
\end{figure}

\subsection{Problem Definition of FL \& PFL}
FL as a disruptive machine learning paradigm has attracted significant attention from researchers and industry in recent years (Figure \ref{fig:intro}, top). This paradigm involves training models on devices locally, mitigating privacy risks associated with transmitting private data to central servers, and providing a viable solution for collaborative learning among multiple clients. 
\par
We assume that each client has its own local dataset, denoted as $\mathcal{D}^{(i)} = (x, y)$, where $x$ and $y$ are the model's input and the expected output, respectively. The collective dataset across all clients is denoted as $\mathcal{D} = \left\{\mathcal{D}^{(i)}\right\}_{i=1}^{N}$, where $N$ denoted the number of clients.
\par
Traditional FL trained the model $f(x;\theta^{(i)})$ on client $i$ with parameter $\theta^{(i)}$. The training objective can be formulated as:
\begin{equation}
    \begin{array}{cc}
         &  \min\limits_{\theta} \mathcal{L}^{\theta} := \frac{1}{N} \sum\limits_{i=1}^{N} \frac{1}{|\mathcal{D}^{(i)}|} \sum\limits_{x \in \mathcal{D}^{(i)}} \mathcal{L}(f(x; \theta^{(i)}), y) \\
    \end{array}
\end{equation}

where the $\theta = \left\{\theta^{(1)}, \theta^{(2)}, \dots, \theta^{(N)} \right\}$, the $\mathcal{L}(f(x; \theta^{(i)}), y)$ denoted the loss between the $f(x; \theta^{(i)})$ and $y$ on client $i$. 
\par
In contrast to traditional FL, research in PFL (Figure \ref{fig:intro}, bottom) aims to address two principal challenges. The first challenge is the poor convergence of heterogeneous data, particularly when learning on \textit{non-IID} data. The accuracy of traditional methods, such as FedAvg, is significantly reduced when training on such data. This performance degradation is attributed to the phenomenon of client drift resulting from local training and synchronization on \textit{non-IID} data. The second challenge is the lack of personalized solutions. In a traditional FL setting, a single global shared model is trained to fit the "\textit{average client}". Consequently, for local distributions that differ significantly from the global distribution, the shared model struggles to generalize effectively. This lack of personalization hinders generalization, especially in practical applications dealing with commonly encountered \textit{non-IID} local data sets. In such cases, a single model is often insufficient.
\par
To address the aforementioned challenges, PFL divides each client's knowledge into two distinct categories: general knowledge and client-specific knowledge. In \textit{non-IID} scenarios, the data distributions of different clients are unique. PFL decomposes the model trained on client data into a shared part, $\theta_{s}$, and a personalized part, $\theta_p$, which is dedicated to learning general knowledge and client-specific knowledge, respectively. Formally, the training objective of the personalized part can be formulated as:
\begin{equation}
    \begin{array}{cc}
         &  \mathop{\min}\limits_{\theta_{p}} \mathcal{L}^{\theta_{p}} := \frac{1}{N} \sum\limits_{i=1}^{N} \frac{1}{|\mathcal{D}^{(i)}|} \sum\limits_{x \in \mathcal{D}^{(i)}} \mathcal{L} (f(x; \theta^{(i)}_{p}), y)  \\
    \end{array}
\end{equation}
where $\theta_{p} = \left\{\theta_{p}^{(1)}, \theta_{p}^{(2)}, \dots,\theta_{p}^{(N)} \right\}$. Consequently, the PFL, which is designed to adapt to statistical heterogeneity, aims to achieve high accuracy across different types and degrees of heterogeneous data, all without prior knowledge of the data distributions.

\subsection{Low-rank Adaptation}
As the number of LLM parameters increases from millions to billions, fine-tuning models by freezing a portion of the parameters becomes increasingly resource-intensive, making it cost-prohibitive. To tackle this challenge, researchers have recently proposed the Parameter-Efficient Fine-Tuning (PEFT) method \cite{hu2023llm}. The objective of PEFT is to enhance the performance of pre-trained LLMs on downstream tasks by reducing the number of fine-tuned parameters and computational complexity, thereby reducing training costs. Consequently, even when constrained by limited computational resources, the LLMs can adapt to downstream tasks, enabling efficient transfer learning.

\begin{figure}
    \centering
    \includegraphics[scale=0.4]{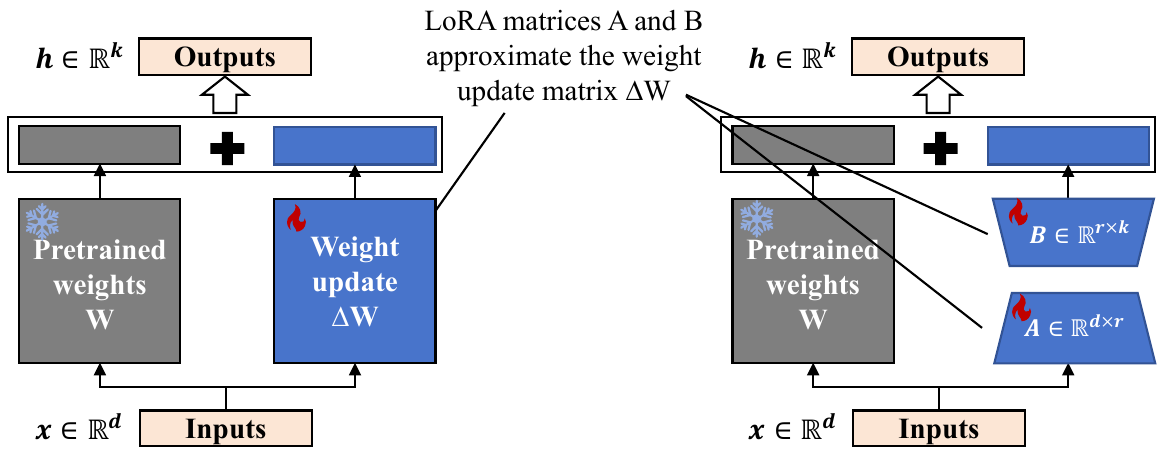}
    \Description[Short description]{Long description of the image.}
    \caption{Comparison of principle between regular fine-tuning (left) and LoRA (right).}
    \label{fig:lora}
\end{figure}
\par
LoRA is currently the most popular PEFT method due to its superior performance compared to other techniques, such as adapter, prefix tuning, and fine-tuning. LoRA embeds domain-specific knowledge into low-rank parameter matrices without altering the original parameters (Figure \ref{fig:lora}). Given a parameter matrix $W_{0} \in \mathbb{R}^{d \times k}$, the low-rank decomposition can be utilized to represent parameter updates $\Delta W$ as follows:
\begin{equation}
    W_{0}  + \Delta W = W_{0} + BA
\end{equation}
where $B \in \mathbb{R}^{d \times r}, A \in \mathbb{R}^{r \times k}$ and $r << \min(d, k)$. $W_{0}$ is freeze in training stage, while $A$ and $ B$ are trainable.  
\par
By learning rank decomposition matrices while simultaneously freezing the original parameters, LoRA reduces the number of trainable parameters significantly (see Figure \ref{fig:lora_parameter}). This substantially diminishes the storage requirements of LLMs tailored for specific tasks and facilitates efficient task switching during deployment without introducing inference latency.

\section{Method} \label{sec:overview}
\subsection{Problem Formulation}
The goal of this paper is to train the LoRA module for diverse clients incorporating the PFL technique to improve performance while maintaining data privacy and efficiency. Building upon this goal, we re-investigated the concept of the PFL in the LLMs era. In particular, we consider that knowledge should be divided into three parts: \textit{basic knowledge}, \textit{domain-general knowledge}, and \textit{personalized knowledge}. \textit{Basic knowledge} is the knowledge that LLM learns from a large corpus, which is not aimed at specific tasks but forms the basis of learning. \textit{Domain-general knowledge} is shared by one or more tasks based on data in a specific domain. \textit{Personalized knowledge }is learned from data distribution specific to each client (\textit{non-IID} scenarios). 
\par
Based on the above concepts, we propose a novel approach, a.k.a \our{}, in the context of PFL. Similar to PFL, we build a LoRA module as the shared part $\theta_{s}$ to learn domain-general knowledge from different clients. Moreover, another LoRA module is built as the personalized part $\theta_{p}^{(i)}$ to learn personalized knowledge for client $i$. Note that the basic knowledge in LLM has not changed, but this part (denoted $\Theta$) is extremely important as the basis of learning. Finally, a function $\mathcal{G}(\cdot)$ is adopted to dynamically fusion \textit{basic knowledge}, \textit{domain-general knowledge}, and \textit{personalized knowledge}. The objective of \our{} is to minimize the sum of the loss of all clients' models, i.e.,

\begin{equation}
    \begin{array}{cc}
         &  \min \limits_{\theta_{s}, \theta_{p}, \mathbf{w}} \frac{1}{N} \sum \limits_{i=1}^{N} \frac{1}{|\mathcal{D}^{(i)}|} \sum \limits_{x \in \mathcal{D}^{(i)}} \mathcal{L} \left(\mathcal{G} (\mathbb{F} (x; \theta_{p}^{(i)}, \theta_{s}, \Theta); \mathbf{w}^{(i)}) \right) \\
    \end{array}
\end{equation}

where $\mathbf{w} = \left\{\mathbf{w}^{(i)}\right\}_{i=1}^{N}$ is the learnable parameters of $\mathcal{G}$ of client $i$, $\mathbb{F} = \left\{f(x;\theta_{p}), f(x; \theta_{s}), f(x;  \Theta)\right\}$ is the mapping of \textit{personalized knowledge} part, \textit{domain-general knowledge} part, and \textit{basic knowledge} part. We will subsequently elaborate on the design of the \our{} .

\begin{figure*}[!tb]
    \centering
    \includegraphics[width=0.9\linewidth]{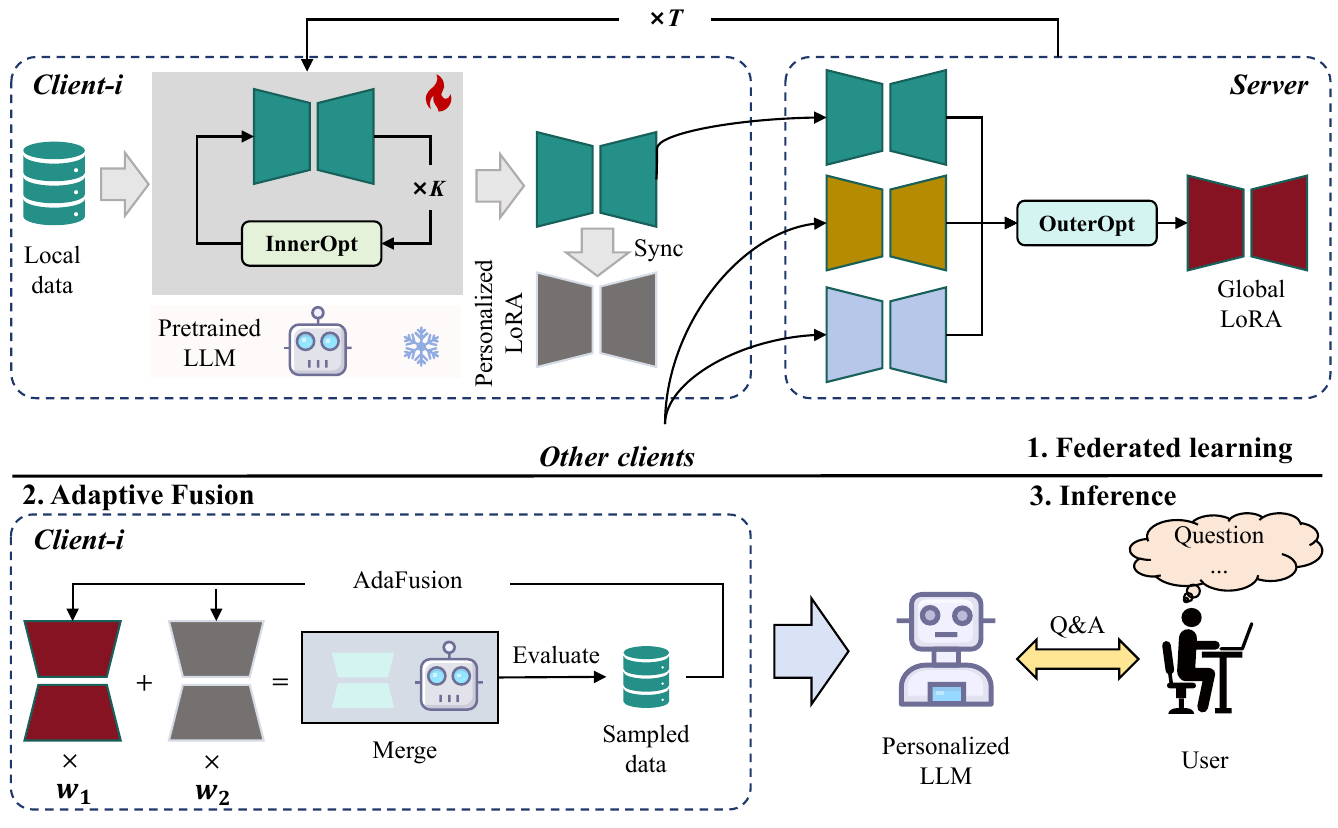}
    \Description[Short description]{Long description of the image.}
    \caption{Framework overview of \our{}.}
    \label{fig:framework}
\end{figure*}

\subsection{Framework Overview}
We introduce the overview of the \our{} framework in this subsection. As shown in Figure \ref{fig:framework}, each client is comprised of four main components: \textbf{LLM}, \textbf{Personalized LoRA module}, \textbf{Global LoRA module,} and \textbf{Adaptive Fusion module}. The personalized LoRA primarily focuses on acquiring knowledge from local data and remains uninvolved in the federated learning process. The global LoRA aggregates knowledge from diverse clients. Specifically, each client uploads its global LoRA parameters to the central server. Then, by employing an external optimizer updates these parameters and subsequently broadcasts them to all clients, thereby completing a communication round of the federated learning process. Finally, the global LoRA parameters and personalized LoRA parameters are aggregated via the AdaFusion module. We will introduce the elaboration in the following subsections. 
\par

\begin{algorithm}[!t]
    \SetAlgoLined 
    \caption{\our{} Training Algorithm} 
    \label{algo:one}
    \KwIn{$\mathcal{D}$, dataset across all clients; $N$, total number of clients; $T$, total number of rounds; $\Theta, \theta_p, \theta_s$, initial LLM, personalized, global LoRA module; $\mathbf{w} = \{\mathbf{w^{1}}, \dots, \mathbf{w^{N}}\}$, parameters of gate function; $H$, the asynchronous update frequency. } 
    \KwOut{$\tilde{S} = f(\cdot; \Theta)$}
    \tcp{Stage 1. Local learning on $N$ clients} 
    \For{$i = 1$ to $N$}{
        \tcp{Training personalized LoRA module via supervised fine-tuning}
        \For{1 to n\_batches}{
            $\mathcal{B} \leftarrow$ Randomly sample $b$ data from $\mathcal{D}^{(i)}$ \\
            $\theta^{(i)}_{p} \leftarrow SFT(f(B; (\Theta, \theta^{(i)}_{p})))$  \\
        }
    }
    \tcp{Stage 2. Federate learning} 
    $\theta_{s}^{(0)} \leftarrow \frac{1}{N}\sum_{i=1}^{N} \theta^{(i)}_{p}$ \tcp*[r]{Initial global LoRA module}
    \For{outer step t = 1 to $T$}{
        \tcp{Update the personalized LoRA module every $H$ communication rounds}  
        \textit{is\_sync} $\leftarrow t \% H$ \\
        \For{client i = 1,\dots,N}{
            $\theta_{s}^{(i)(t)} \leftarrow \theta_{s}^{(t-1)}$ \\
            \tcp{Inner optimization}
            $\theta_{s}^{(i)(t)} \leftarrow \text{InnerOpt}(\theta_{s}^{(i)(t)}, \mathcal{D}^{(i)}, K)$ \\
            \If {\textit{is\_sync}}{
                \tcp{Synchronize personalized and global LoRA module}
                $\theta^{(i)}_{p} \leftarrow \theta_{s}^{(i)(t)}$ 
            }
        }
        \tcp{Averaging outer gradients}
        $\Delta^{(t)} \leftarrow \frac{1}{N} \sum_{i=1}^{N} (\theta_{s}^{(t-1)} - \theta_{s}^{(i)(t)})$ \\
        \tcp{Outer optimization}
        $\theta_{s}^{(t)} \leftarrow \text{OuterOpt}(\theta_{s}^{(t-1)}, \Delta^{(t)})$ \\
    }
    \tcp{Stage 3. Adaptive fusion function} 
    \For{client i = 1, \dots, N}{
        $\mathbf{w}^{(i)}$ $\leftarrow$ FusionOpt($\theta_{p}^{(i)}, \mathbf{w^{(i)}}$) \\
    }
    
    \textbf{return} $\theta_{s}, \theta_{p}, \mathbf{w}$
\end{algorithm}

\subsection{Local Learning}
\par
First, we start from the basic model, which can be initialized with pretrained LLM, such as LLaMA2 \cite{touvron2023llama} and ChatGLM \cite{du2022glm, zeng2022glm}. We assume that there are $N$ clients, each capable of training a model replica, and $N$ shards of data with one for each worker. Then, we initialize personalized LoRA with parameters $\theta_p$, and utilize SFT to capture knowledge from local private data (\textit{Algorithm} \ref{algo:one}, \textit{line 1-6}). We formulate this process in each client as follows:

\begin{equation}
     \theta^{(i)}_{p} \leftarrow SFT(f(B; (\Theta, \theta^{(i)}_{p})))
\end{equation}

where $1 \leq i \leq N$ and $\mathcal{B}$ is randomly sample batch data from $D^{(i)}$.
\par
Finally, the average of all clients $\theta_p^{(i)}$ is used to initialize the global LoRA parameters $\theta_s$ for the subsequent federated learning process (in line 7), which could reduce the computational cost in the first round. This process is formulated as follows:
\begin{equation}
    \theta_{s}^{(0)} \leftarrow \frac{1}{N}\sum_{i=1}^{N} \theta^{(i)}_{p}
\end{equation}

\subsection{Federate Learning}
At this stage, there are two optimization processes. There is an outer optimization, which consists of $T$ outer steps. At each outer step $t$, gradients from each client are gathered, averaged, and sent to an outer optimizer (OuterOpt) to update the global LoRA module parameters (\textit{Algorithm} \ref{algo:one}, \textit{line 17-18}). Afterward, the global LoRA module parameters are re-dispatched to each client (\textit{Algorithm} \ref{algo:one}, \textit{line 11}).
\par
In each phase, each client performs its own inner optimization for $K$ steps independently and in parallel using an inner optimizer, denoted by InnerOpt. Each client samples data from local data and updates its personalized LoRA module parameters (\textit{Algorithm} \ref{algo:one}, \textit{line 12}). It should be noted that the inner optimization consists of $K > 1$ steps; for instance, $K=3$ is the default setting in our experiments. Therefore, communication across clients is minimal.
\par
Once all clients have completed their inner optimization step, the cumulative gradients of each client are averaged. The resulting outer gradient is then utilized to update the global LoRA parameters, serving as the starting point for the subsequent round of inner optimizations. Communication among workers is only necessary during this phase, occurring once every $K$ inner optimization steps. In total, a client undergoes $T \times K$ inner steps during this stage.
\par
Specifically, the inner optimizer (InnerOpt) is used as AdamW \cite{zhuang2022understanding}, which is a variant of the Adam optimizer that separates weight decay from the gradient update based on the observation that the weight decay formulation is different when applied to SGD and Adam. As for the outer optimizer (OuterOpt), we use Nesterov momentum \cite{sutskever2013importance}, because it gave the best convergence in previous works \cite{douillard2023diloco}. When OuterOpt is set to SGD, the outer optimizer is equivalent to vanilla FedAVG \cite{mcmahan2017communication}. If the total number of outer optimization steps ($T$) is set to 1, then our approach reduces to 'souping' \cite{wortsman2021learning}. Finally, if the number of inner optimization steps ($K$) is set to 1 and InnerOpt is set to SGD, our approach is equivalent to large-batch training with data parallelism.

\par
Overall, iterating the above process $T$ times, all clients contain a personalized LoRA and a global LoRA. The personalized LoRA contains the knowledge learned from the local data, and the global LoRA contains the knowledge of all clients and is aligned with the local knowledge by SFT. The degree of alignment can be adjusted by the inner optimization steps $K$. Increasing $K$ will give more attention to local knowledge. This process ensures that, as the federated learning progresses through multiple iterations, each client's global LoRA parameters are enriched with a harmonious blend of both local and global knowledge, contributing to a comprehensive and collaborative model refinement.

\subsection{Adaptive Fusion}
In this stage, we implement an element-wise method, a.k.a. AdaFusion, to adaptively combine both the personalized LoRA module and the global LoRA module. This process integrates the corresponding parameters of the LoRA modules, requiring the modules being combined to have the same rank $r$ to properly align the structures. For the client $i$, assuming that $m_{i} = A_i B_i$ and $i=1$ or $2$, the combined LoRA module $\hat{m}$ can be formulated as follows:

\begin{equation}
    \hat{m} = (w_{1}A_{1} + w_{2}A_{2}) (w_{1}B_{1} + w_{2}B_{2})
\end{equation}

The objective is to modify the coefficients $\mathbf{w^{(i)}} = \{w_1^{(i)}, w_2^{(i)}\}$ to enhance the performance on each client. To achieve this, a fusion optimizer (FusionOpt) is introduced. An intuitive approach is to implement the FusionOpt using gradient-based optimization methods, such as standard gradient descent. However, this approach necessitates the construction of a hypernetwork on each client, which can be challenging due to the substantial GPU memory and time requirements \cite{huang2023lorahub}. This is a significant challenge for heterogeneous clients. We observe that the number of optimized parameters is relatively small (only $w_1$ and $w_2$ in each client). Consequently, we opt for gradient-free methods for optimization rather than gradient descent.
\par
Inspired by previous work \cite{huang2023lorahub}, we employ a black-box optimization technique to find the optimal $\mathbf{w}$. This optimization process hinges on minimizing the cross-entropy loss, with the objective of determining the optimal coefficients $w_1$ and $w_2$  that minimize the loss $\mathcal{L}_{c}$ across a set of few-shot examples denoted as $Q$. Furthermore, we employ $L_1$ regularization, which penalizes extreme values by imposing a constraint on the sum of the absolute values of $\mathbf{w}$. Consequently, the objective function is formulated as follows, which minimizes the combined loss and the regularization term.
\begin{equation}
   \min_{\mathbf{w}} \mathcal{L} = \mathcal{L}_{c} + \lambda^{(i)} |\mathbf{w^{(i)}}| 
\end{equation}
where $\lambda$ is the hyperparameter.

\section{Evaluation} \label{sec:evaluation}
In this section, we conducted a series of experiments on both log-based anomaly detection and medical diagnosis scenarios to answer the following research questions:
\begin{itemize}
    \item \textbf{RQ1:} How does \our{} perform compared to baselines under different federation setups?
    \item \textbf{RQ2:} How does the communication round affect accuracy?
    \item \textbf{RQ3:} How does the communication frequency affect accuracy?
    \item \textbf{RQ4:} How does the asynchronous update frequency affect accuracy?
    \item \textbf{RQ5:} How effective are combine both personalized LoRA and global LoRA?
    \item \textbf{RQ6:} How to balance communication and computation costs?
    \item \textbf{RQ7:} What are the advantages of the AdaFusion compared with other implementations?
\end{itemize}

\subsection{Experiment Setup}
\textbf{Dataset.} To evaluate our approach and baselines on different scenarios. We select two scenarios including both system log analysis and medical diagnosis. Specifically, we utilize three public log datasets from LogHub \cite{zhu2023loghub}: BGL \cite{oliner2007supercomputers}, Spirit \cite{stearley2008bad}, and Thunderbird \cite{oliner2007supercomputers} in the system log analysis.  For medical diagnosis, we employ the bio-medicine dataset \footnote{https://huggingface.co/datasets/AdaptLLM/medicine-tasks}. The statistic of data in both scenarios is present in Table \ref{tab:dataset}. 
\par
\textbf{Characteristics.} In \textit{Scenario-1}, three datasets exhibit varying sample sizes, ranging from 10,000 samples each. The input length distributions within these datasets vary significantly, indicating diverse data characteristics and potential challenges in handling inputs of different lengths. In \textit{Scenario-2}, five datasets are utilized, with the number of samples ranging from 500 to 2,000. Similarly, these datasets display diverse input length distributions, reflecting the heterogeneity present in federated learning settings. The analysis demonstrates the necessity of addressing the diverse data characteristics present in federated learning scenarios to achieve robust model performance.

\par
\textbf{Data Preprocessing.} In log analysis (\textit{Scenario-1}), our primary focus is on the task of log-based anomaly detection. First, the unstructured logs within each dataset are formatted using the log parsing method. Then, the formatted logs are sorted based on timestamps and sequences are generated using fixed-size sliding windows (default is 50). Following the existing work \cite{qi2023loggpt}, we retain only the log content sequence, which excludes information such as timestamps. Finally, a conversation template (see appendix) is constructed to guide the LLM in determining whether the log content sequence is an anomaly.
In bio-medicine (\textit{Scenario-2}), we refer to the templates for instructions and conversation prompts in \cite{cheng2024adapting}.

\begin{table}[!tb]
\centering
\caption{Statistic of dataset used in the experiments.}
\label{tab:dataset}
\begin{tabular}{llll} 
\toprule
Scenario                    & Dataset     & \#Samples & \begin{tabular}[c]{@{}l@{}}Input length \\distribution\end{tabular}  \\ 
\midrule
\multirow{3}{*}{\textit{Scenario-1}} & BGL         & 10,000   & 78\textasciitilde{}5,478                         \\
                            & Spirt       & 10,000   & 78\textasciitilde{}1,628                         \\
                            & Thunderbird & 10,000   & 145\textasciitilde{}1,828                         \\ 
\midrule
\multirow{5}{*}{\textit{Scenario-2}} & ChemProt    & 500       & 2,690\textasciitilde{}4,690                \\
                            & MPQ         & 610       & 816\textasciitilde{}2,050                         \\
                            & PubMedQA    & 1,000     & 413\textasciitilde{}2,940  \\
                            & RCT         & 2,000     & 1,810\textasciitilde{}3,530                          \\
                            & USMLE       & 1,270     & 73\textasciitilde{}3,560                          \\
\bottomrule
\end{tabular}
\end{table}

\par
Moreover, we employ the Dirichlet \textit{non-IID} setting, a popular utilized framework in contemporary FL research \cite{lin2020ensemble, ma2022layer}. Within this framework, the data for each participating client is generated from a Dirichlet distribution denoted as $Dir(\alpha)$. The parameter $\alpha$ governs the shape of the distribution; as it increases, the degree of class imbalance within each client's dataset diminishes. Consequently, the Dirichlet \textit{non-IID} setting enables us to evaluate the performance of our methodologies across a diverse array of \textit{non-IID} scenarios, thereby providing insights into their robustness and adaptability. Unless otherwise specified, the default value for $\alpha$ is assumed to be $0.5$ in the following experiments. 
\par
Finally, each client's local data is further divided into training and testing sets in a ratio of 8:2. Consequently, the testing set is stored locally by each client, following the same distribution as the local training set.
\par
\textbf{Baseline.} To verify the effectiveness of \our{}, we compare it with six baselines. The details of each algorithm are as follows:
\begin{itemize}
    \item \textit{Local:}  Each client’s local LLM was trained with its local training set independently.  
    \item \textit{FedAVG \cite{mcmahan2017communication}} was a popular FL algorithm that aggregates model parameters by weighted averaging. The basic idea of FedAvg is to upload the parameters of local models to the server, where the server computes the average of all model parameters and sends this average back to all local devices. This process can be iterated many times until convergence is achieved.
    \item \textit{FedKD \cite{wu2022communication}} was a communication-efficient federated learning method, that utilizes adaptive mutual knowledge distillation and dynamic gradient compression.
    \item \textit{FedAMP \cite{huang2021personalized}} introduced federated attentive message passing to foster collaborations among clients with similar data in federated learning. Convergence proofs for convex and non-convex models are established. Additionally, a heuristic method enhances FedAMP's performance with deep neural networks. 
    \item \textit{FedRep \cite{collins2021exploiting}} presented a novel federated learning framework, emphasizing shared data representation and unique client-specific heads, achieving efficient convergence and superior performance in heterogeneous federated environments.
    \item \textit{FedRoD \cite{chen2021bridging}} introduced a novel federated learning framework, Federated Robust Decoupling, which addressed the challenge of balancing generic and personalized model performance. By decoupling the model's dual duties into two prediction tasks and leveraging robust loss functions, FedRoD achieved state-of-the-art performance. 
\end{itemize}

\par
\begin{figure}
    \centering
    \includegraphics[scale=0.4]{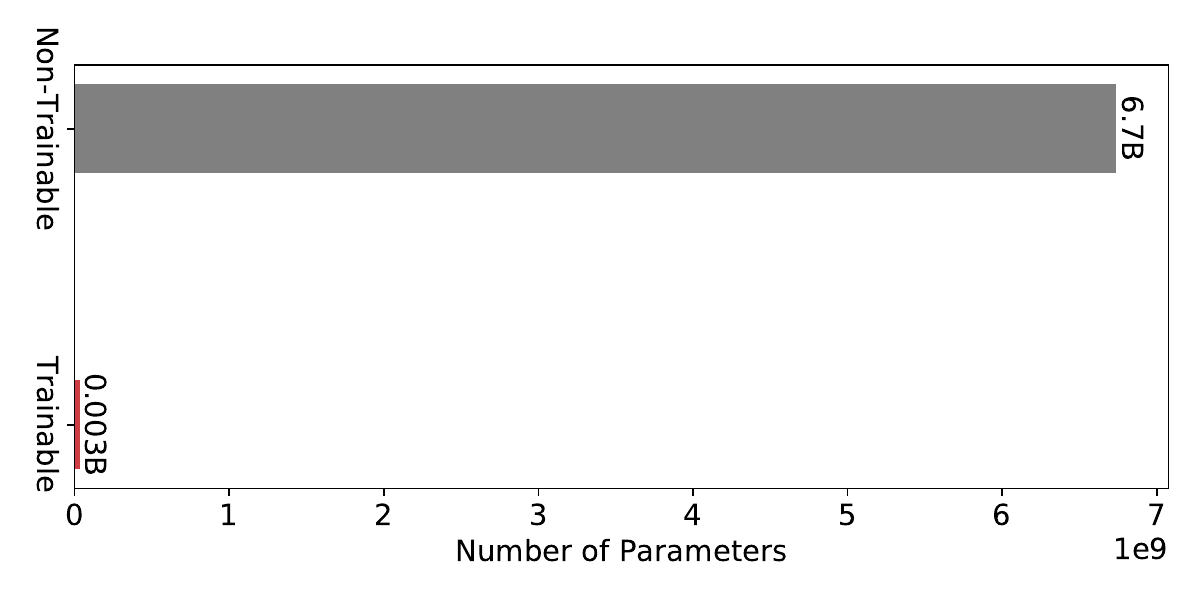}
    \caption{Trainable Parameters vs Non-Trainable Parameters.}
    \label{fig:lora_parameter}
\end{figure}
\textbf{Implementation.} We have chosen LLaMA2-7B \cite{touvron2023llama} as the backbone of LLM for our evaluation. LLaMA2-7B is an encoder-only architecture developed by Meta and has been pre-trained on over 1 trillion tokens. Despite having a smaller parameter scale (7 billion parameters), it has demonstrated superior performance on various benchmarks, even outperforming GPT-3 (175 billion parameters) \cite{floridi2020gpt}. LLaMA2-7B may be a suitable choice for researchers and startups with resource constraints due to its compact parameter size. Then, we implement LoRA \footnote{https://huggingface.co/docs/peft/index} to significantly decrease computational and storage costs, while yielding performance comparable to a fully fine-tuned model. As shown in Figure \ref{fig:lora_parameter}, we compared the number of trainable parameters and the total number of parameters. The trainable parameters are only about $0.03B$, which is only $\sim 5\%$ of the total parameters. Finally, we fine-tuning the LLaMA2-7B with \our{} framework. The details of the settings are as follows:
\par
For the inner optimization, each client's local LLM was fine-tuned by LoRA with its local training set for 3 epochs with the learning rate $lr=0.0002$, batch size $b = 1$. The optimizer is PagedAdamW32bit \footnote{https://huggingface.co/docs/bitsandbytes/main/reference/optim/adamw} with the momentum $m=0.9$, learning rate decay $\eta=0.1$, and weight decay $\gamma=0.01$.  For the outer optimization, we set the learning rate $lr=0.001$, momentum $m=0.5$, and the total communication across all clients $T=30$. For the adaptive fusion function, we set the max inference step is 5, and the regularization factor $\lambda^{(i)} = 0.05$ in client $i$ .

\par
\textbf{Evaluation Metrics.} For the \textit{Scenario-1}, we use the \textbf{F1 score} to evaluate the accuracy of the log-based anomaly detection task. For the \textit{Scenario-2}, we compare the output to the groundtruth answer, and the accuracy of an individual data example is considered true if the model output matches the groundtruth. To facilitate clarity and consistency in our descriptions, we will use the term \textbf{accuracy} uniformly in the subsequent sections.

\subsection{RQ1: Performance on different scenarios}
\textbf{Experiment Settings.} The performance of \our{} is compared with baseline methods on four datasets. We set the client number $N=5$, the local update epochs $K=3$, the batch size $B=1$, and the total communication round $T=30$. Each client is assigned an equal number of samples, and the training and testing sets are split with a ratio of 8:2 for each client. The results are present in Table \ref{tab:baseline}.
\par

\our{} emerges as the standout performer, consistently exhibiting the highest mean accuracy across both scenarios and varying values of $\alpha$. This indicates \our{}'s robustness and effectiveness in mitigating the challenges associated with \textit{non-IID} data distribution. The comparative analysis also reveals interesting trends regarding the impact of class imbalance on model performance. As $\alpha$ increases, denoting reduced class imbalance, all methods demonstrate improved performance, suggesting that models find it easier to learn with more balanced class distributions. However, the extent of improvement varies among methods, with \our{} consistently outperforming others. These findings underscore the importance of employing tailored federated learning approaches like \our{} to optimize model performance in scenarios characterized by \textit{non-IID} data distributions. Furthermore, the reported standard deviations highlight the variability in performance across multiple runs, emphasizing the need for robustness in federated learning algorithms to ensure consistent results in real-world applications. 
\par
In summary, the \our{} offers several advantages, including high performance, global collaboration capabilities, model robustness, and data privacy protection. These benefits make \our{} a powerful solution for FL problems and promote better cooperation among all parties involved.

\begin{table*}
\centering
\caption{Comparison results under Dirichlet \textit{non-IID} on both \textit{Scenario-1} and \textit{Scenario-2}.}
\label{tab:baseline}
\begin{tabular}{l!{\vrule width \lightrulewidth}ccc!{\vrule width \lightrulewidth}ccc} 
\toprule
                                        & \multicolumn{3}{c!{\vrule width \lightrulewidth}}{\textit{Scenario-1}}                                                                                                                & \multicolumn{3}{c}{\textit{Scenario-2}}                                                                                                                 \\ 
\midrule
\diagbox{Method}{$\alpha$} & \multicolumn{1}{c}{$0.1$} & \multicolumn{1}{c}{$0.5$} & \multicolumn{1}{c!{\vrule width \lightrulewidth}}{$1.0$} & \multicolumn{1}{c}{$0.1$} & \multicolumn{1}{c}{$0.5$} & \multicolumn{1}{c}{$1.0$}  \\ 
\midrule
Local     & 61.30 $\pm$ 2.01     & 59.07 $\pm$ 1.35    & 57.50 $\pm$ 0.61      & 56.30 $\pm$ 0.83     & 53.80 $\pm$ 0.68       & 50.80 $\pm$ 0.59                                         \\
FedAVG    & 44.30 $\pm$ 1.07   & 44.53 $\pm$ 0.77   & 44.10 $\pm$ 0.43   & 38.90 $\pm$ 0.32           & 38.70 $\pm$ 0.27      & 36.80 $\pm$ 0.19                \\
FedKD        & 63.30 $\pm$ 1.11   & 62.57 $\pm$ 0.82     & 61.60 $\pm$ 0.38       & 49.40  $\pm$ 0.93     & 47.10  $\pm$ 0.74   & 46.70  $\pm$ 0.72                                         \\
FedAMP    & 73.90  $\pm$ 1.81   & 70.20  $\pm$ 1.25   & 69.80 $\pm$ 0.89   & 50.10  $\pm$ 2.33  & 47.80  $\pm$ 1.91  & 46.60  $\pm$ 1.76                                         \\
FedRep     & 73.70 $\pm$ 1.67   & 72.87 $\pm$ 0.83   & 65.90 $\pm$ 0.71     & 48.10 $\pm$ 0.32  & 47.30  $\pm$ 0.34  & 42.40  $\pm$ 0.46                                         \\
FedRoD    & 75.60 $\pm$ 2.86   & 74.93 $\pm$ 2.05  & 71.30 $\pm$ 2.12  & 50.20 $\pm$ 1.71   & 49.20 $\pm$ 0.86   & 46.20  $\pm$ 0.78                                          \\
\midrule
\our{}    & \textbf{76.30 $\pm$ 1.89 }    & \textbf{78.17 $\pm$ 1.11}     & \textbf{75.60 $\pm$ 0.47}   & \textbf{57.70 $\pm$ 0.68}  & \textbf{56.20 $\pm$ 0.83}    & \textbf{55.50 $\pm$ 0.72}                                          \\
\bottomrule
\end{tabular}
\end{table*}

\begin{figure*}[!t]
    \centering
    \subfigure[N=3]{\includegraphics[scale=0.28]{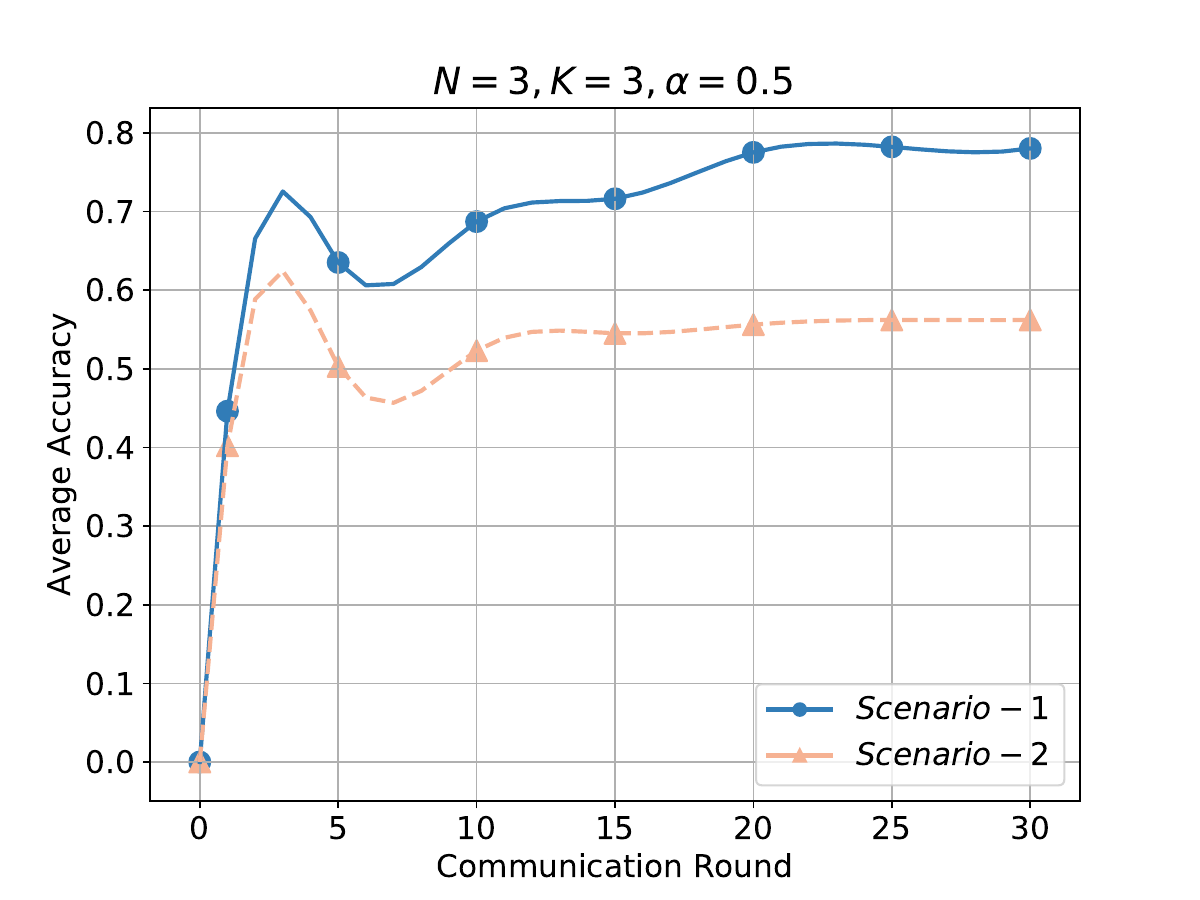}}
    \subfigure[N=5]{\includegraphics[scale=0.28]{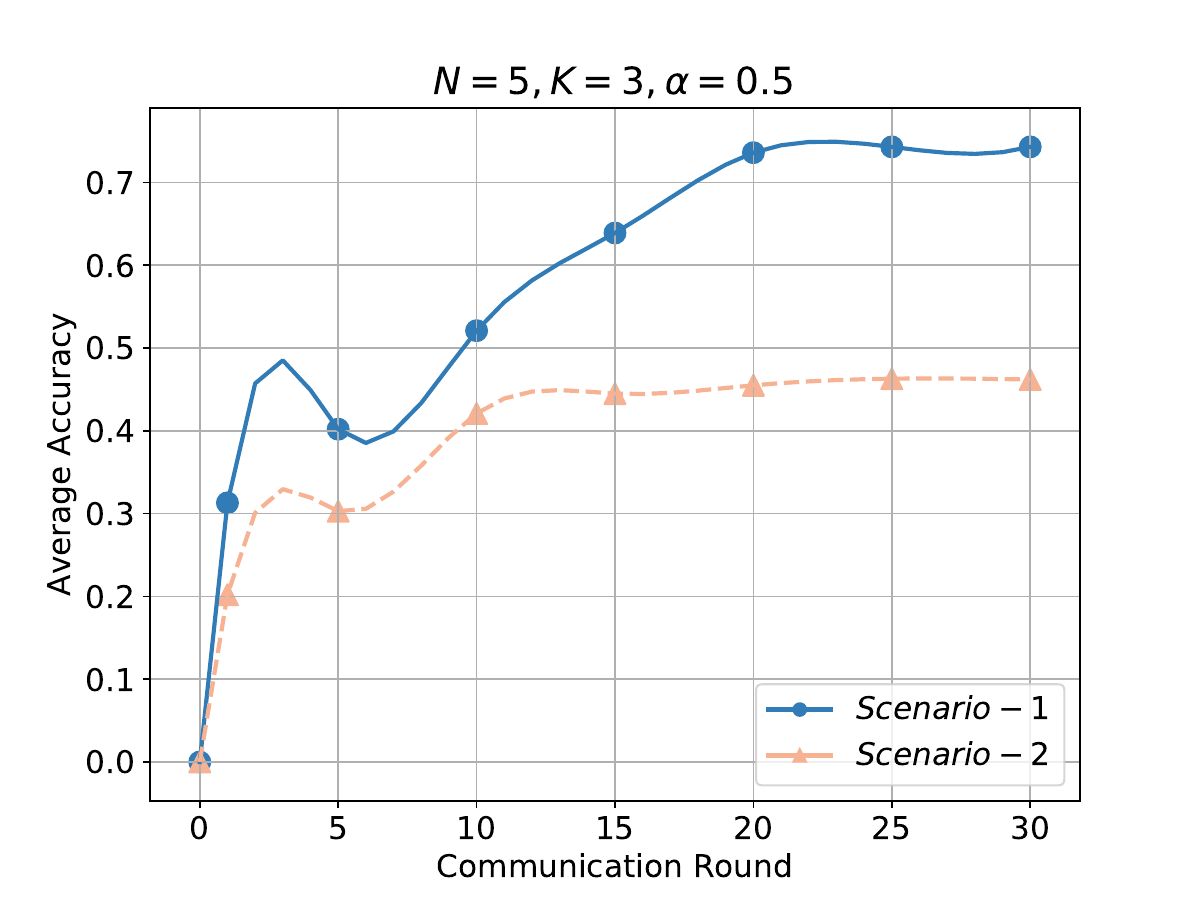}}
    \subfigure[N=10]{\includegraphics[scale=0.28]{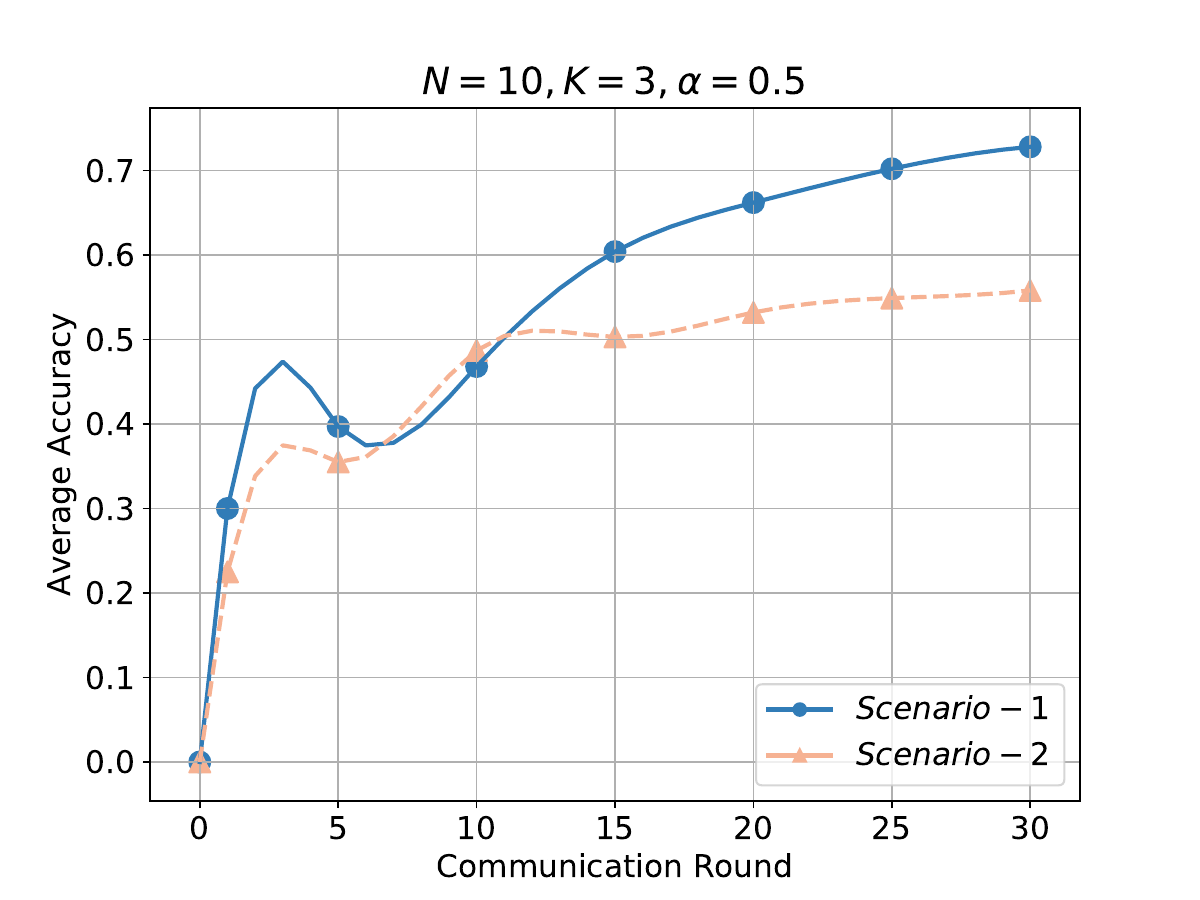}}
    \Description[Short description]{Long description of the image.}
    \caption{The average accuracy varies with the communication round for different settings of the number of clients $N={3, 5, 10}$ and InnerOpt step $K=3$.}
    \label{fig:communication}
\end{figure*}
\subsection{RQ2: Performance with varies communication round}
\textbf{Experiment Settings.} To evaluate the performance of varying communication rounds ($T$), we increase $T$ from 0 to 30. The number of clients is set to $N$=3, 5 and 7, with a fixed number of local update iterations ($K=3$), and the batch size $B=1$. The results are shown in Figure \ref{fig:communication}.
\par
It's clear that as the number of communication rounds increases, there is a consistent improvement in average accuracy across different scenarios, particularly when considering different numbers of clients. This consistency in performance regardless of the number of clients suggests that the effect of communication rounds on accuracy is robust and generalizable across different scales of client populations.
For example, when comparing the results across different values of $N$ (e.g. $N=3$, $N=5$, $N=10$), we observe that in each case there is a noticeable trend of improving average accuracy as the number of communication rounds increases. This suggests that the collaborative nature of PFL, facilitated by increased communication rounds, consistently leads to improved model performance, regardless of the number of participating clients.
\par
This consistency highlights the importance of communication rounds as a fundamental aspect of PFL, enabling effective coordination and aggregation of decentralized model updates across a wide range of client populations. It suggests that, regardless of the specific deployment scenario or the size of the FL system, increasing the number of communication rounds can generally lead to improved model convergence and accuracy, thus emphasizing its importance in the FL paradigm.

\begin{figure}[!t]
    \centering
    \subfigure[Scenario-1]{\includegraphics[scale=0.16]{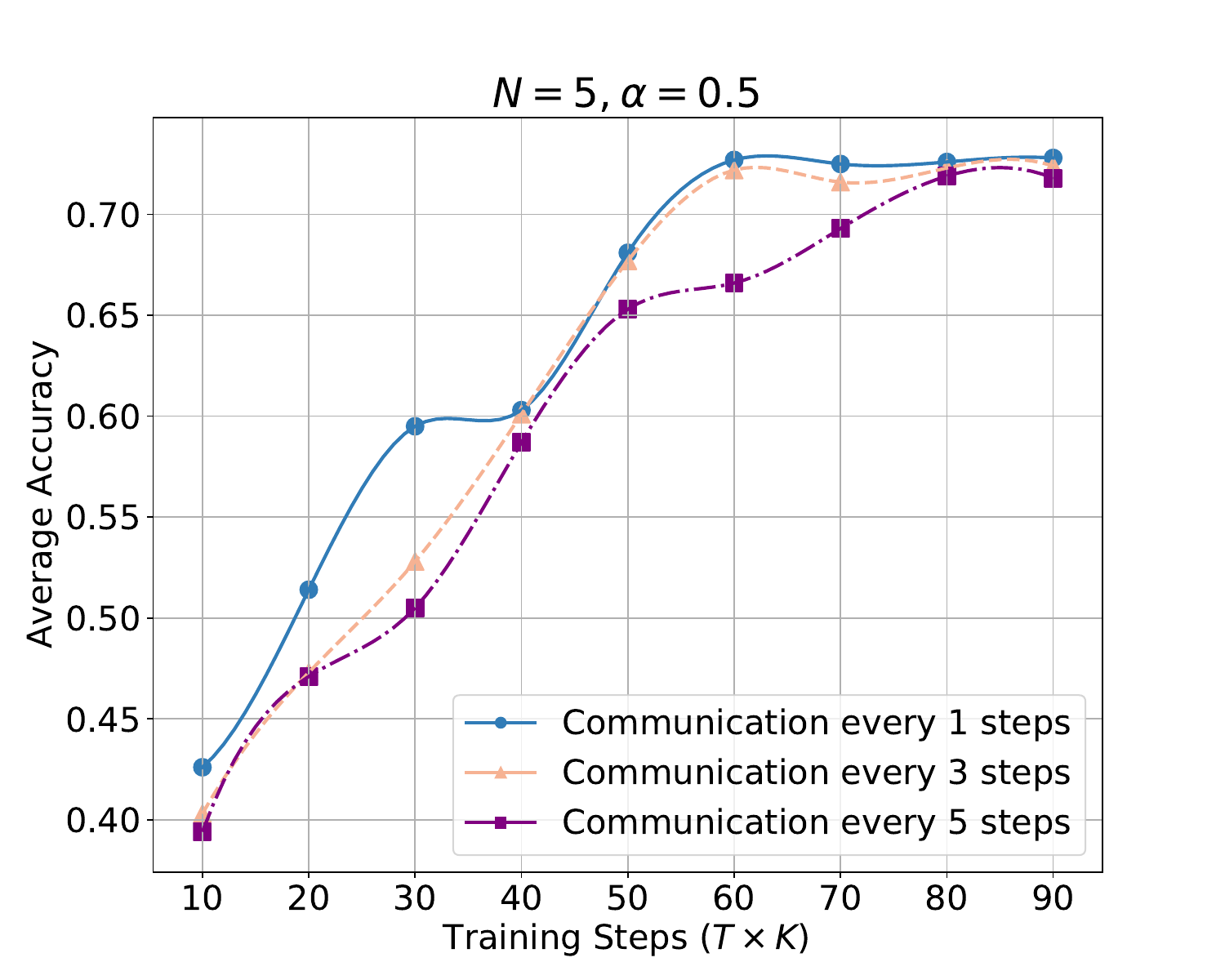}}
    \subfigure[Scenario-2]{\includegraphics[scale=0.16]{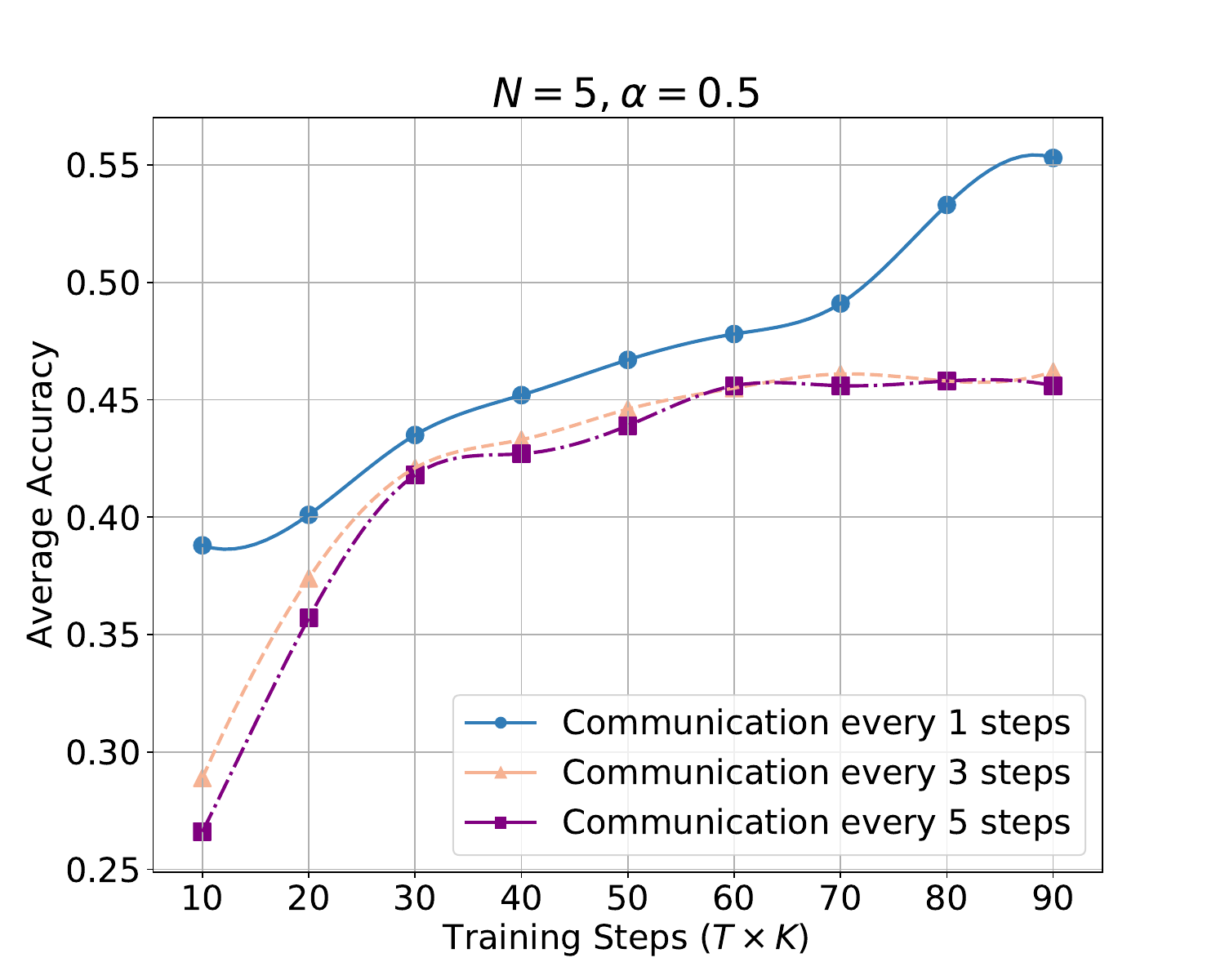}}
    \Description[Short description]{Long description of the image.}
    \caption{Varying the communication frequency every $K= \{1, 3, 5\}$ steps.}
    \label{fig:communication_frequency}
\end{figure}

\subsection{RQ3: Performance with varies communication frequency}
\textbf{Experiment Settings.} In practical scenarios, the communication overhead between clients can be significant, influenced by factors such as regional network conditions and heterogeneous hardware. Doing communication and aggregation operations at the end of each training iteration (i.e., when $K=1$) may be the sub-optimal choice. However, existing methods often consider communication every $K\leq20$ step. This setting may be too frequent for fine-tuning LLMs via LORA, considering the powerful learning capabilities of LLMs and the limited scale of local data. To determine an appropriate communication frequency, we conducted experiments with different InnerOpt update frequencies, specifically $K$=1, 3, and 5. The results are shown in Figure \ref{fig:communication_frequency}.
\par
The results show notable trends in the performance of the \our{} under different InnerOpt update frequency (personalized step) and OuterOpt update frequency (communication round) settings. When comparing InnerOpt update frequencies ($K$), both scenarios show a trend of improved average accuracy as the InnerOpt update frequency decreases. This suggests that incorporating more frequent communication, such as changing from $K=5$ (in \textcolor[RGB]{128,0,128}{purple}) to $K=1$ (in \textcolor{blue}{blue}), helps to improve LLM convergence and ultimately leads to better performance and the importance of more aggregation through communication in the training process. However, it's worth noting that the rate of improvement can vary depending on the specific scenario and the frequency of updates.
\par
The results highlight the importance of adjusting both the update frequency of InnerOpt and OuterOpt in PFL systems. While more frequent communication generally leads to improved LLM convergence and accuracy, there is a delicate balance to be struck to reduce computational and communication overhead without compromising performance, particularly in scenarios with limited client resources or network conditions.

\begin{figure}[!tb]
    \centering
    \includegraphics[scale=0.25]{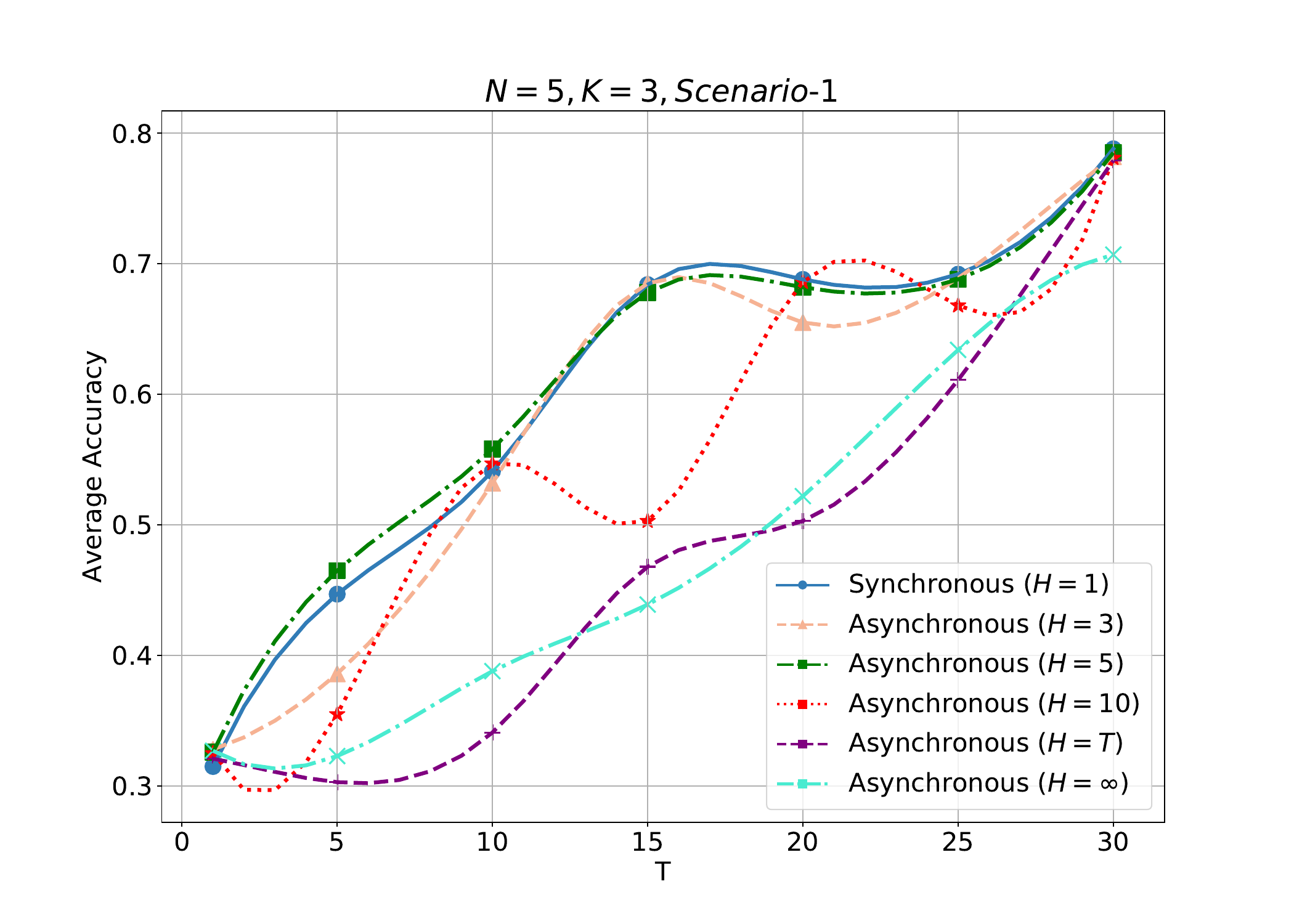}
    \Description[Short description]{Long description of the image.}
    \caption{Impact of different synchronous update frequency with $\alpha=0.5$. Note that when $H=1$ the asynchronous equivalent to synchronous, and when $H=\infty$ the personalized LoRA module is frozen after local learning (Stage 1).}
    \label{fig:sync_frequency}
\end{figure}

\subsection{RQ4: Performance with varies synchronous update frequency}
\textbf{Experiment Settings.} In \our{}, all clients perform supervised fine-tuning after receiving aggregated global parameters from the server (global LoRA). Subsequently, the client has the option to choose whether to synchronize the parameters with the personalized LoRA. To investigate the impact of different rounds of parameter synchronization on performance, we conducted experiments by varying the synchronous update frequency, where $H=\{1, 3, 5, 10, T, \infty\}$,  and observed the resulting performance changes. The experimental results are shown in Figure \ref{fig:sync_frequency}.
\par
The experimental results provide insights into the impact of different parameter synchronization frequencies on performance. First, in the synchronous (in \textcolor{blue}{blue}) scenario, where parameters are synchronized between the personalized LoRA and the global LoRA after each InnerOpt step, we observe a stable and gradual improvement in accuracy with increasing communication round. This indicates that frequent parameter synchronization helps to maintain LLM alignment across clients, contributing to consistent performance improvements. In contrast, asynchronous synchronization scenarios ($H>1$) show different performance trends. Asynchronous updates with H=3 and $H=5$ show fluctuating accuracy, suggesting that moderate synchronization frequencies may not always result in significant performance gains. However, asynchronous synchronization with $H=10$ (in \textcolor{red}{red}) and $H=T$ (in \textcolor[RGB]{128,0,128}{purple}) shows more consistent improvements in accuracy over time, suggesting that less frequent but more substantial parameter synchronization may be beneficial. Notably, in the asynchronous update scenario with $H=\infty$ (in \textcolor[RGB]{73,235,208}{teal}), where parameters are not synchronized, we observe a decrease in accuracy compared to other synchronization frequencies, highlighting the importance of parameter synchronization in maintaining LLM performance.
\par
In summary, these results suggest that striking a balance between synchronization frequency and performance gains is essential for optimizing \our{} performance. Furthermore, the decrease in accuracy observed in the absence of parameter synchronization ($H=\infty$) underscores the need for parameter synchronization to maintain model alignment and performance across distributed clients.

\begin{table}[!tb]
\centering
\caption{The effect of both Personalized and Global LoRA module with $\alpha = 0.5$.}
\label{tab:lora_ablation}
\begin{tabular}{ccc} 
\toprule
Methods    & \textit{Scenario-1}   & \textit{Scenario-2}  \\ 
\midrule
LLaMA2-7B (\textit{0-shot})    & 16.13 & 31.36        \\
LLaMA2-7B (\textit{5-shot})    & 23.51 & 32.69        \\
Personalized (\textit{standalone})  & 58.26 & 47.61        \\
Global (\textit{standalone}) & 61.30 & 56.30        \\
\midrule
\our{(Ours)} & 78.17 & 56.20 \\
\bottomrule
\end{tabular}
\end{table}

\subsection{RQ5: Effectiveness of combining personalized LoRA and global LoRA}
\textbf{Experiment Settings.}  To validate the effectiveness of combining the personalized LoRA module with the global LoRA module, we conducted ablation experiments comparing it with the standalone personalized LoRA module and the standalone global LoRA module. We set the number of clients to $N=5$, the number of communication rounds to $T=30$, and the synchronization frequency to $H=T$. Additionally, comparisons were made with standard LLMs under \textit{0-shot} and \textit{5-shot} settings. The results are shown in Table \ref{tab:lora_ablation}
\par
We can see that the standalone personalized and global LoRA modules show significant improvements over the off-the-shelf model with both 0-shot and 5-shot settings. In particular, the personalized module achieves significantly higher accuracy than the off-the-shelf model in both \textit{Scenario-1} and \textit{Scenario-2}, indicating its effectiveness in adapting to non-IDD data. Similarly, the global module also shows notable improvements over the off-the-shelf model, suggesting its ability to capture overarching patterns or preferences across the dataset. Notably, the global module outperforms the personalized module in \textit{Scenario-1}, while the personalized module performs slightly better in \textit{Scenario-2}. This discrepancy may indicate that the nature of the data or the characteristics of the users in each scenario affects the effectiveness of the modules differently. Furthermore, the combination of both personalized and global modules in the proposed approach yields the highest scores in both scenarios, suggesting that leveraging both individualized and general patterns leads to improved performance.
\par
In summary, the experimental results suggest that both personalized and global LoRA modules contribute significantly to improving performance compared to the off-the-shelf model. The personalized module excels in capturing personalized knowledge in each client, while the global module effectively captures general patterns across clients. The combination of both modules in the proposed approach leads to superior performance, indicating the importance of leveraging both personalized and general patterns in practice. This highlights the potential of hybrid approaches that integrate personalized and global perspectives to achieve better results in different scenarios.

\begin{table*}[!tb]
\centering
\caption{Comparison of different batch size increasing strategies ($\alpha=0.5$).}
\label{tab:various_bs}
\begin{tabular}{lccccc} 
\toprule
\multirow{2}{*}{Method}                       & \multirow{2}{*}{Communication} & \multirow{2}{*}{Time} & \multirow{2}{*}{Compute  Data} & \multicolumn{2}{c}{Average Accuarcy}  \\
\cmidrule{5-6}
                                              &                                &                       &                                & \textit{Scenario-1} & \textit{Scenario-2 }      \\
\midrule
Baseline                                      & 0                              & $1\times$                    & $1\times$                             & 78.18      & 57.95            \\
Baseline, $4\times$ batch size with data parallelism & $4\times N$            & $1\times$                    & $4\times$                            & 77.74      & 56.48            \\
Baseline, $4\times$batch size with
  microbatching   & 0                              & $4\times$                    & $4\times$                            & 77.72      & 56.69            \\
Baseline, $4\times$update                            & 0                              & $4\times$                   & $4\times$                             & 73.55      & 55.03            \\
\our{}                                     & $4\times \frac{T}{K}$                          & $1\times$                    & $4\times$                             & 77.39      & 55.83            \\
\bottomrule
\end{tabular}
\end{table*}

\subsection{RQ6: Trade-off between communication cost and computational cost}
\textbf{Experiment Settings.} The baseline methods involve variations in communication, time, and compute/data usage. Specifically, the first baseline utilizes standard settings with a single GPU for both data processing and model training, while the second employs data parallelism with four GPUs. The third baseline increases the batch size per GPU by four, resulting in slower training times, while the fourth keeps the batch size constant but increases the accumulation factor to 4. The results are present in Table \ref{tab:various_bs}.
\par
In \textit{Scenario-1}, the baseline method employing data parallelism achieves optimal performance, benefiting from efficient communication and compute/data utilization. However, approaches with increased batch size or accumulation factor suffer from longer training times and lower average accuracy. This suggests that while parallelizing data across multiple GPUs can enhance performance, overly increasing batch size or accumulation factor may introduce bottlenecks in computation and negatively impact model convergence.
\par
In \textit{Scenario-2}, larger batch sizes outperform smaller ones in the baseline methods, indicating the potential benefits of aggregating more data for training. However, an increase in the accumulation factor leads to performance degradation, highlighting the importance of balancing compute resources with training efficiency. These findings underscore the complex trade-offs involved in federated learning, where optimizing communication and computation strategies is essential for achieving optimal model performance across diverse scenarios.
\par
The results demonstrate the intricate interplay between communication, computation, and model performance in federated learning. While certain strategies, such as data parallelism or larger batch sizes, may excel in specific scenarios, they may not generalize well across different settings. \our{} emerges as a promising approach, striking a balance between communication and computation costs while maintaining competitive performance across diverse scenarios, including log anomaly detection and medical question answering.

\begin{table*}[!tb]
\centering
\caption{Comparison with different fusion methods ($N=5$).}
\label{tab:fusion_impl}
\begin{tabular}{l!{\vrule width \lightrulewidth}ccc!{\vrule width \lightrulewidth}ccc} 
\toprule
\multirow{2}{*}{\diagbox{Fusion Method}{$\alpha$}} & \multicolumn{3}{c!{\vrule width \lightrulewidth}}{\textit{Scenario-1}}                                              & \multicolumn{3}{c}{\textit{Scenario-2}}                                                                              \\ 
\cmidrule{2-7}
                                                                    & 0.1                               & 0.5                                & 1                                 & 0.1                               & 0.5                                & 1                                  \\ 
\midrule
Random                                                              & 32.88 $\pm$ 5.81 & 36.33 $\pm$ 5.46  & 41.61 $\pm$ 4.81 & 48.15 $\pm$ 2.97 & 48.73 $\pm$ 3.37~ & 48.43 $\pm$ 1.35  \\
Average                                                             & 36.94 $\pm$ 2.66 & 46.57 $\pm$ 2.87  & 43.21 $\pm$ 1.90 & 39.53 $\pm$ 1.77 & 39.77 $\pm$ 1.58~ & 43.28 $\pm$ 0.85  \\
Sum                                                                 & 68.47 $\pm$ 1.76 & 67.86 $\pm$ 2.05~ & 64.81 $\pm$ 1.84 & 48.76 $\pm$ 1.05 & 47.36 $\pm$ 1.31~ & \textbf{49.18 $\pm$ 1.11}  \\
\midrule
AdaFusion   & \textbf{76.30 $\pm$ 1.89} & \textbf{78.17 $\pm$ 1.11}  & \textbf{75.60 $\pm$ 0.47} & \textbf{57.70 $\pm$ 0.68} & \textbf{56.20 $\pm$ 0.83}  & 48.50 $\pm$ 0.72  \\
\bottomrule
\end{tabular}
\end{table*}

\subsection{RQ7: Advantages of AdaFusion method}
\textbf{Experiment Settings.} To evaluate the advantages of the \textit{AdaFusion} method over other fusion methods, we compared it with three baseline methods: \textit{1)}. \textit{Random}, where weights between 0 and 1 are randomly assigned to the personalized LoRA ($w_1$) and the global LoRA ($w_2$); \textit{2)}. \textit{Average}, where the parameters of the two LoRA modules are arithmetically averaged (equivalent to $w_1=0. 5$ and $w_2=0.5$); \textit{3)}. \textit{Sum}, where the parameters of the two LoRA modules are added directly (equivalent to $w_1=1.0$ and $w_2=1.0$). We evaluated the results under different \textit{non-IID} settings (e.g., $\alpha=\{0.1, 0.5, 1.0\}$), as shown in Table \ref{tab:fusion_impl}.
\par
In \textit{Scenario-1}, \textit{AdaFusion} consistently outperforms the other fusion methods across all $\alpha$ values. This indicates that \textit{AdaFusion} effectively adapts to the varying degrees of data heterogeneity, achieving superior average accuracy compared to Random, Average, and Sum fusion methods. Notably, the robust performance of \textit{AdaFusion} underscores its ability to dynamically adjust fusion weights based on the characteristics of each client's data distribution, leading to enhanced model convergence and predictive accuracy.
\par
However, the performance of \textit{AdaFusion} in \textit{Scenario-2} exhibits slight fluctuations across different $\alpha$ values. While \textit{AdaFusion} generally maintains competitive accuracy at $\alpha= \{0.1, 0.5\}$,  it is inferior to the Sum fusion method at $\alpha=1.0$. This discrepancy suggests that the efficacy of fusion methods may vary depending on the specific characteristics of the FL environment, such as the degree of data heterogeneity and the complexity of the learning task. Furthermore, it highlights the importance of considering the interplay between fusion methods and the inherent properties of the data distribution when designing FL algorithms.
\par
In summary, the results reveal the nuanced performance of fusion methods in FL scenarios with varying degrees of data heterogeneity. AdaFusion emerges as a robust fusion approach, demonstrating superior adaptability and performance in \textit{Scenario-1} across all $\alpha$ values. Its ability to dynamically adjust fusion weights based on local data characteristics enhances model convergence and predictive accuracy, making it well-suited for addressing the challenges of \textit{non-IID} data distributions in FL. However, further investigation is warranted to understand the factors influencing the performance of fusion methods in different scenarios and to optimize their efficacy for diverse FL applications.


\section{Related Work} \label{sec:related}
In this section, we briefly review the related work of personalized federate learning and parameter-efficient fine-tuning.
\par
\textit{Personalized federate learning (PFL).} The primary objective of PFL is to address the challenge encountered in conventional federated learning, where the global model lacks personalization for local tasks or datasets, leading to slow convergence and suboptimal performance in heterogeneous data (\textit{non-IID}) scenarios. PFL tackles this challenge through two key approaches \cite{tan2022towards}: global model personalization and learning personalized model.
\par
\textbf{Global model personalization}. To address the challenge of FL training on \textit{non-IID} data, the process of personalizing a global model globally can be divided into two stages. Firstly, a shared global model is trained via FL. Secondly, additional training is performed on local data to adapt to personalization. It is important to note that this process can be further divided into data-based and model-based methods. 
\par
The data-based method addresses the client drift problem, effectively converting \textit{non-IID} into IID settings. Thus, the unresolved problem (\textit{non-IID}) can be transferred to the resolved problem (IID) setting so that existing methods \cite{mcmahan2017communication} can process it. Research has investigated the use of data augmentation methods to enhance statistical homogeneity. However, applying existing over-sampling and under-sampling methods directly to FL settings, where client data is distributed and private, is challenging. To overcome this, researchers have developed data augmentation techniques tailored for FL settings \cite{hao2021towards}.
\par
Data-based approaches in FL mitigate client drift but often alter local data distributions, potentially losing valuable insights into client behavior diversity. Model-based global model personalization FL approaches aim to address this by either learning a robust global FL model for future client personalization or enhancing local model adaptation performance. These methods prioritize leveraging client behavior diversity to improve model personalization and adaptation, ensuring more effective and flexible FL outcomes. Researchers mainly use regularization \cite{acar2021federated}, meta-learning \cite{fallah2020personalized}, and transfer learning \cite{chen2020fedhealth} to associate global models and local models for solution personalization. For instance, FedProx \cite{li2020federated} addresses statistical and systems heterogeneity. It introduced a proximal term to ensure the aggregation of incompletely calculated partial information and provide theoretical convergence guarantees under non-identical data distributions and system-level constraints. pFedMe \cite{t2020personalized} leverages Moreau envelopes to regularize clients' loss functions, separating personalized optimization from global learning. Theoretically, pFedMe achieves state-of-the-art convergence rates, with quadratic speedup for strongly convex and 2/3-order sublinear speedup for smooth nonconvex objectives. FedHealth \cite{chen2020fedhealth} is the first federated transfer learning framework for wearable healthcare, which aggregates data via federated learning and tailors personalized models through transfer learning, ensuring accuracy and privacy. 
\par
\textbf{Learning personalized model.} Unlike global model personalization, the learning personalized model has only a single training stage to achieve the effect of model personalization. Furthermore, it can be divided into structure-based methods and similarity-based methods. 
\par
Architecture-based PFL approaches customize model designs for individual clients. Techniques such as parameter decoupling \cite{jin2022accelerated} incorporate personalization layers for each client, while knowledge distillation methods \cite{zhu2021data} enable personalized model architectures for individual clients. Tailoring the architecture to each client's specific requirements improves the effectiveness and adaptability of federated learning systems, resulting in improved model personalization. SplitFed \cite{thapa2022splitfed} merges FL and Split Learning (SL) to mitigate their drawbacks. SplitFed incorporates a refined architecture with differential privacy and PixelDP for enhanced data privacy and model robustness. He et al. proposed FedGKT \cite{he2020group}, a group knowledge transfer training algorithm. FedGKT employs an alternating minimization approach to train small CNNs on edge nodes and periodically transfer their knowledge via knowledge distillation to a large server-side CNN. FedGKT reduced edge computation demand, lower communication bandwidth for large CNNs, and asynchronous training, while maintaining accuracy comparable to FedAvg. 
\par
Similarity-based approaches focus on modeling client relationships to achieve personalization. These approaches aim to learn personalized models for each client, with related clients learning similar models \cite{ouyang2021clusterfl}. Various types of client relationships are explored in PFL research. Multi-task learning \cite{hu2021private} and model interpolation methods \cite{chen2023fedsoup} consider pairwise client relationships, while clustering techniques \cite{briggs2020federated} examine group-level client relationships. These approaches leverage client similarities to enhance model personalization and performance in FL scenarios.  Huang et al. proposed FedAMP \cite{huang2021personalized}, which employs federated attentive message passing to facilitate collaborations among similar clients in federated learning with \textit{non-IID} data. Convergence is established for both convex and non-convex models. Additionally, a heuristic method is introduced to enhance performance for clients using deep neural networks. HeteroFL \cite{diao2020heterofl} is tailored for heterogeneous clients with varying computation and communication capabilities, which allows for training heterogeneous local models with different complexities while producing a single global inference model. It challenges the assumption of uniform model architectures across clients and demonstrates strategies to enhance federated learning training efficiency. Specifically, it achieves efficiency by adaptively distributing subnetworks based on clients' capabilities. FedGroup \cite{duan2020fedgroup} has introduced a new data-driven distance measure and a newcomer device cold start mechanism to improve scalability. The optimization process is divided into subgroups using the Kmeans++ algorithm, which can be combined with the FL optimizer FedProx to achieve even greater improvements. 

\textit{Parameter-efficient fine-tuning (PEFT).} With the exponential increase in the number of parameters in state-of-the-art LLMs, fine-tuning for each downstream task has become exceedingly costly in terms of both time and resources. Xu et al. \cite{xu2023parameter} aims to address this challenge by adapting LLMs to new tasks through the updating of only a small number of parameters, potentially including new ones. Several significant PEFT approaches have emerged to achieve this goal.
\begin{itemize}
    \item \textbf {Adapter-based methods} \cite{kim2023efficient, cai2022fedadapter} allow efficient fine-tuning of parameters by adding a small Adapter layer within the Transformer block. Each Adapter layer contains a lower projection matrix, an activation function, and an upper projection matrix. The lower projection matrix maps the input features to the bottleneck dimension, the activation function is used for non-linear transformation, and the upper projection matrix maps the bottleneck features back to the original dimensions. In this way, the adapter can introduce additional parameters to adapt to downstream tasks, while keeping most parameters unchanged.
    \item \textbf{Soft prompt} are learnable tensors concatenated with the input embeddings, which can be optimized to a dataset. However, one disadvantage is that they are not human-readable, as they do not correspond directly to embeddings of actual words. The soft prompt methods encompass several approaches, including prompt tuning \cite{jia2022visual}, prefix tuning \cite{vos2022towards}, P-tuning \cite{liu2021p}, and multi-task prompt tuning \cite{asai2022attempt}.
    \item \textbf{Reparameterized-based methods} adds learnable parameters during the training process by constructing a low-rank representation of the original model parameters to achieve efficient fine-tuning of parameters. Such methods include Low-Rank Adaptation (LoRA) \cite{hu2021lora}, LoRA Derivatives \cite{yang2024moral}, and Hybrid PEFT \cite{runwal2024peft}.
\end{itemize}
\par
\textbf{Compared with the previous works:} Our work belongs to the category of reparameterized-based methods, but can also be extended to other PEFT methods. Most related to our work is the work of \cite{yi2023fedlora, wu2024fedlora}, who combined the ideas of LoRA and iteratively trained the target model. However, \our{} is specifically designed for practical scenarios where there is a scarcity of labeled data on each client and aims to minimize the number of fine-tuned parameters. In contrast to vanilla LoRA, our method carefully combines personalized LoRA and global LoRA, leveraging a fusion optimizer to integrate diverse knowledge from each client. We have developed a novel PFL framework that enables the training of both personalized and global LoRA, effectively harnessing the collaborative potential across clients. This strategy enhances the overall knowledge utilization and performance. Finally, \our{} offers a practical solution for federated learning scenarios with limited labeled data on each client, while efficiently incorporating both personalized and global knowledge.

\section{Conclusion} \label{sec:conclusion}
In this paper, we have presented \our{}, a novel variant of personalized federated learning framework based on LoRA tuning, which overcomes the lack of labeled data and the gap between the high cost and suboptimal of existing FL-based frameworks in the popular LLM scenario. \our{} utilized dual LoRA modules to capture personalized and global knowledge, effectively leveraging collaboration across all clients and reducing the number of trainable parameters. Moreover, it employed an adaptive fusion algorithm to combine the parameters of two LoRAs to improve the performance. It achieves state-of-the-art in both log analysis and medical diagnosis scenarios demonstrating stability and robustness with respect to \textit{non-IID} degree, the number of clients, and communication rounds and frequency. Finally, \our{} outperforms other increase batch size strategies with respect to communication cost, time spent training and the amount of compute \& data used.
\par
In our future work, we will explore the application of \our{} to more real-world scenarios and evaluate its effectiveness and practicality in industrial environments, such as natural language understanding, sentiment analysis, and recommendation systems. Moreover, we will continue to enhance the robustness and efficiency of \our{} by exploring advanced techniques for model optimization and communication compression.

\bibliographystyle{ACM-Reference-Format}
\bibliography{ref}

@inproceedings{xu2024unilog,
  title={UniLog: Automatic Logging via LLM and In-Context Learning},
  author={Xu, Junjielong and Cui, Ziang and Zhao, Yuan and Zhang, Xu and He, Shilin and He, Pinjia and Li, Liqun and Kang, Yu and Lin, Qingwei and Dang, Yingnong and others},
  booktitle={Proceedings of the 46th IEEE/ACM International Conference on Software Engineering},
  pages={1--12},
  year={2024}
}

@article{bai2024federated,
  title={Federated Fine-tuning of Large Language Models under Heterogeneous Language Tasks and Client Resources},
  author={Bai, Jiamu and Chen, Daoyuan and Qian, Bingchen and Yao, Liuyi and Li, Yaliang},
  journal={arXiv preprint arXiv:2402.11505},
  year={2024}
}

@inproceedings{zhang2024towards,
  title={Towards building the federatedGPT: Federated instruction tuning},
  author={Zhang, Jianyi and Vahidian, Saeed and Kuo, Martin and Li, Chunyuan and Zhang, Ruiyi and Yu, Tong and Wang, Guoyin and Chen, Yiran},
  booktitle={ICASSP 2024-2024 IEEE International Conference on Acoustics, Speech and Signal Processing (ICASSP)},
  pages={6915--6919},
  year={2024},
  organization={IEEE}
}

@misc{
wu2024fedlora,
title={FedLo{RA}: When Personalized Federated Learning Meets Low-Rank Adaptation},
author={Xinghao Wu and Xuefeng Liu and Jianwei Niu and Haolin Wang and Shaojie Tang and Guogang Zhu},
year={2024},
url={https://openreview.net/forum?id=bZh06ptG9r}
}

@article{sheller2020federated,
  title={Federated learning in medicine: facilitating multi-institutional collaborations without sharing patient data},
  author={Sheller, Micah J and Edwards, Brandon and Reina, G Anthony and Martin, Jason and Pati, Sarthak and Kotrotsou, Aikaterini and Milchenko, Mikhail and Xu, Weilin and Marcus, Daniel and Colen, Rivka R and others},
  journal={Scientific reports},
  volume={10},
  number={1},
  pages={12598},
  year={2020},
  publisher={Nature Publishing Group UK London}
}

@article{runwal2024peft,
  title={From PEFT to DEFT: Parameter Efficient Finetuning for Reducing Activation Density in Transformers},
  author={Runwal, Bharat and Pedapati, Tejaswini and Chen, Pin-Yu},
  journal={arXiv preprint arXiv:2402.01911},
  year={2024}
}

@article{yang2024moral,
  title={MoRAL: MoE Augmented LoRA for LLMs' Lifelong Learning},
  author={Yang, Shu and Ali, Muhammad Asif and Wang, Cheng-Long and Hu, Lijie and Wang, Di},
  journal={arXiv preprint arXiv:2402.11260},
  year={2024}
}

@article{asai2022attempt,
  title={ATTEMPT: Parameter-efficient multi-task tuning via attentional mixtures of soft prompts},
  author={Asai, Akari and Salehi, Mohammadreza and Peters, Matthew E and Hajishirzi, Hannaneh},
  journal={arXiv preprint arXiv:2205.11961},
  year={2022}
}

@article{liu2021p,
  title={P-tuning v2: Prompt tuning can be comparable to fine-tuning universally across scales and tasks},
  author={Liu, Xiao and Ji, Kaixuan and Fu, Yicheng and Tam, Weng Lam and Du, Zhengxiao and Yang, Zhilin and Tang, Jie},
  journal={arXiv preprint arXiv:2110.07602},
  year={2021}
}

@inproceedings{vos2022towards,
  title={Towards parameter-efficient automation of data wrangling tasks with prefix-tuning},
  author={Vos, David and D{\"o}hmen, Till and Schelter, Sebastian},
  booktitle={NeurIPS 2022 First Table Representation Workshop},
  year={2022}
}

@inproceedings{jia2022visual,
  title={Visual prompt tuning},
  author={Jia, Menglin and Tang, Luming and Chen, Bor-Chun and Cardie, Claire and Belongie, Serge and Hariharan, Bharath and Lim, Ser-Nam},
  booktitle={European Conference on Computer Vision},
  pages={709--727},
  year={2022},
  organization={Springer}
}

@article{cai2022fedadapter,
  title={Fedadapter: Efficient federated learning for modern nlp},
  author={Cai, Dongqi and Wu, Yaozong and Wang, Shangguang and Lin, Felix Xiaozhu and Xu, Mengwei},
  journal={arXiv preprint arXiv:2205.10162},
  year={2022}
}

@article{kim2023efficient,
  title={Efficient Federated Learning with Pre-Trained Large Language Model Using Several Adapter Mechanisms},
  author={Kim, Gyunyeop and Yoo, Joon and Kang, Sangwoo},
  journal={Mathematics},
  volume={11},
  number={21},
  pages={4479},
  year={2023},
  publisher={MDPI}
}

@inproceedings{briggs2020federated,
  title={Federated learning with hierarchical clustering of local updates to improve training on non-IID data},
  author={Briggs, Christopher and Fan, Zhong and Andras, Peter},
  booktitle={2020 International Joint Conference on Neural Networks (IJCNN)},
  pages={1--9},
  year={2020},
  organization={IEEE}
}

@inproceedings{chen2023fedsoup,
  title={FedSoup: Improving Generalization and Personalization in Federated Learning via Selective Model Interpolation},
  author={Chen, Minghui and Jiang, Meirui and Dou, Qi and Wang, Zehua and Li, Xiaoxiao},
  booktitle={International Conference on Medical Image Computing and Computer-Assisted Intervention},
  pages={318--328},
  year={2023},
  organization={Springer}
}

@article{hu2021private,
  title={Private multi-task learning: Formulation and applications to federated learning},
  author={Hu, Shengyuan and Wu, Zhiwei Steven and Smith, Virginia},
  journal={arXiv preprint arXiv:2108.12978},
  year={2021}
}

@inproceedings{ouyang2021clusterfl,
  title={Clusterfl: a similarity-aware federated learning system for human activity recognition},
  author={Ouyang, Xiaomin and Xie, Zhiyuan and Zhou, Jiayu and Huang, Jianwei and Xing, Guoliang},
  booktitle={Proceedings of the 19th Annual International Conference on Mobile Systems, Applications, and Services},
  pages={54--66},
  year={2021}
}

@inproceedings{zhu2021data,
  title={Data-free knowledge distillation for heterogeneous federated learning},
  author={Zhu, Zhuangdi and Hong, Junyuan and Zhou, Jiayu},
  booktitle={International conference on machine learning},
  pages={12878--12889},
  year={2021},
  organization={PMLR}
}

@inproceedings{jin2022accelerated,
  title={Accelerated federated learning with decoupled adaptive optimization},
  author={Jin, Jiayin and Ren, Jiaxiang and Zhou, Yang and Lyu, Lingjuan and Liu, Ji and Dou, Dejing},
  booktitle={International Conference on Machine Learning},
  pages={10298--10322},
  year={2022},
  organization={PMLR}
}

@article{chen2020fedhealth,
  title={Fedhealth: A federated transfer learning framework for wearable healthcare},
  author={Chen, Yiqiang and Qin, Xin and Wang, Jindong and Yu, Chaohui and Gao, Wen},
  journal={IEEE Intelligent Systems},
  volume={35},
  number={4},
  pages={83--93},
  year={2020},
  publisher={IEEE}
}

@article{fallah2020personalized,
  title={Personalized federated learning with theoretical guarantees: A model-agnostic meta-learning approach},
  author={Fallah, Alireza and Mokhtari, Aryan and Ozdaglar, Asuman},
  journal={Advances in Neural Information Processing Systems},
  volume={33},
  pages={3557--3568},
  year={2020}
}

@article{acar2021federated,
  title={Federated learning based on dynamic regularization},
  author={Acar, Durmus Alp Emre and Zhao, Yue and Navarro, Ramon Matas and Mattina, Matthew and Whatmough, Paul N and Saligrama, Venkatesh},
  journal={arXiv preprint arXiv:2111.04263},
  year={2021}
}

@inproceedings{hao2021towards,
  title={Towards fair federated learning with zero-shot data augmentation},
  author={Hao, Weituo and El-Khamy, Mostafa and Lee, Jungwon and Zhang, Jianyi and Liang, Kevin J and Chen, Changyou and Duke, Lawrence Carin},
  booktitle={Proceedings of the IEEE/CVF conference on computer vision and pattern recognition},
  pages={3310--3319},
  year={2021}
}

@article{floridi2020gpt,
  title={GPT-3: Its nature, scope, limits, and consequences},
  author={Floridi, Luciano and Chiriatti, Massimo},
  journal={Minds and Machines},
  volume={30},
  pages={681--694},
  year={2020},
  publisher={Springer}
}

@inproceedings{wortsman2021learning,
  title={Learning neural network subspaces},
  author={Wortsman, Mitchell and Horton, Maxwell C and Guestrin, Carlos and Farhadi, Ali and Rastegari, Mohammad},
  booktitle={International Conference on Machine Learning},
  pages={11217--11227},
  year={2021},
  organization={PMLR}
}

@article{douillard2023diloco,
  title={DiLoCo: Distributed Low-Communication Training of Language Models},
  author={Douillard, Arthur and Feng, Qixuan and Rusu, Andrei A and Chhaparia, Rachita and Donchev, Yani and Kuncoro, Adhiguna and Ranzato, Marc'Aurelio and Szlam, Arthur and Shen, Jiajun},
  journal={arXiv preprint arXiv:2311.08105},
  year={2023}
}

@article{zhuang2022understanding,
  title={Understanding adamw through proximal methods and scale-freeness},
  author={Zhuang, Zhenxun and Liu, Mingrui and Cutkosky, Ashok and Orabona, Francesco},
  journal={Transactions on Machine Learning Research},
  year={2022}
}

@article{wu2023brief,
  title={A brief overview of ChatGPT: The history, status quo and potential future development},
  author={Wu, Tianyu and He, Shizhu and Liu, Jingping and Sun, Siqi and Liu, Kang and Han, Qing-Long and Tang, Yang},
  journal={IEEE/CAA Journal of Automatica Sinica},
  volume={10},
  number={5},
  pages={1122--1136},
  year={2023},
  publisher={IEEE}
}

@inproceedings{ma2022layer,
  title={Layer-wised model aggregation for personalized federated learning},
  author={Ma, Xiaosong and Zhang, Jie and Guo, Song and Xu, Wenchao},
  booktitle={Proceedings of the IEEE/CVF conference on computer vision and pattern recognition},
  pages={10092--10101},
  year={2022}
}

@article{lin2020ensemble,
  title={Ensemble distillation for robust model fusion in federated learning},
  author={Lin, Tao and Kong, Lingjing and Stich, Sebastian U and Jaggi, Martin},
  journal={Advances in Neural Information Processing Systems},
  volume={33},
  pages={2351--2363},
  year={2020}
}

@article{huang2023lorahub,
  title={Lorahub: Efficient cross-task generalization via dynamic lora composition},
  author={Huang, Chengsong and Liu, Qian and Lin, Bill Yuchen and Pang, Tianyu and Du, Chao and Lin, Min},
  journal={arXiv preprint arXiv:2307.13269},
  year={2023}
}

@inproceedings{oliner2007supercomputers,
  title={What supercomputers say: A study of five system logs},
  author={Oliner, Adam and Stearley, Jon},
  booktitle={37th annual IEEE/IFIP international conference on dependable systems and networks (DSN'07)},
  pages={575--584},
  year={2007},
  organization={IEEE}
}

@inproceedings{stearley2008bad,
  title={Bad words: Finding faults in spirit's syslogs},
  author={Stearley, Jon and Oliner, Adam J},
  booktitle={2008 Eighth IEEE International Symposium on Cluster Computing and the Grid (CCGRID)},
  pages={765--770},
  year={2008},
  organization={IEEE}
}

@inproceedings{zhu2023loghub,
  title={Loghub: A large collection of system log datasets for ai-driven log analytics},
  author={Zhu, Jieming and He, Shilin and He, Pinjia and Liu, Jinyang and Lyu, Michael R},
  booktitle={2023 IEEE 34th International Symposium on Software Reliability Engineering (ISSRE)},
  pages={355--366},
  year={2023},
  organization={IEEE}
}

@article{hu2023llm,
  title={Llm-adapters: An adapter family for parameter-efficient fine-tuning of large language models},
  author={Hu, Zhiqiang and Wang, Lei and Lan, Yihuai and Xu, Wanyu and Lim, Ee-Peng and Bing, Lidong and Xu, Xing and Poria, Soujanya and Lee, Roy Ka-Wei},
  journal={arXiv preprint arXiv:2304.01933},
  year={2023}
}

@article{touvron2023llama,
  title={Llama 2: Open foundation and fine-tuned chat models},
  author={Touvron, Hugo and Martin, Louis and Stone, Kevin and Albert, Peter and Almahairi, Amjad and Babaei, Yasmine and Bashlykov, Nikolay and Batra, Soumya and Bhargava, Prajjwal and Bhosale, Shruti and others},
  journal={arXiv preprint arXiv:2307.09288},
  year={2023}
}

@inproceedings{du2022glm,
  title={GLM: General Language Model Pretraining with Autoregressive Blank Infilling},
  author={Du, Zhengxiao and Qian, Yujie and Liu, Xiao and Ding, Ming and Qiu, Jiezhong and Yang, Zhilin and Tang, Jie},
  booktitle={Proceedings of the 60th Annual Meeting of the Association for Computational Linguistics (Volume 1: Long Papers)},
  pages={320--335},
  year={2022}
}

@article{zeng2022glm,
  title={Glm-130b: An open bilingual pre-trained model},
  author={Zeng, Aohan and Liu, Xiao and Du, Zhengxiao and Wang, Zihan and Lai, Hanyu and Ding, Ming and Yang, Zhuoyi and Xu, Yifan and Zheng, Wendi and Xia, Xiao and others},
  journal={arXiv preprint arXiv:2210.02414},
  year={2022}
}

@inproceedings{
cheng2024adapting,
title={Adapting Large Language Models via Reading Comprehension},
author={Daixuan Cheng and Shaohan Huang and Furu Wei},
booktitle={The Twelfth International Conference on Learning Representations},
year={2024},
url={https://openreview.net/forum?id=y886UXPEZ0}
}

@inproceedings{sutskever2013importance,
  title={On the importance of initialization and momentum in deep learning},
  author={Sutskever, Ilya and Martens, James and Dahl, George and Hinton, Geoffrey},
  booktitle={International conference on machine learning},
  pages={1139--1147},
  year={2013},
  organization={PMLR}
}

@inproceedings{qi2023loggpt,
  title={Loggpt: Exploring chatgpt for log-based anomaly detection},
  author={Qi, Jiaxing and Huang, Shaohan and Luan, Zhongzhi and Yang, Shu and Fung, Carol and Yang, Hailong and Qian, Depei and Shang, Jing and Xiao, Zhiwen and Wu, Zhihui},
  booktitle={2023 IEEE International Conference on High Performance Computing \& Communications, Data Science \& Systems, Smart City \& Dependability in Sensor, Cloud \& Big Data Systems \& Application (HPCC/DSS/SmartCity/DependSys)},
  pages={273--280},
  year={2023},
  organization={IEEE}
}

@article{tan2022towards,
  title={Towards personalized federated learning},
  author={Tan, Alysa Ziying and Yu, Han and Cui, Lizhen and Yang, Qiang},
  journal={IEEE Transactions on Neural Networks and Learning Systems},
  year={2022},
  publisher={IEEE}
}

@inproceedings{mcmahan2017communication,
  title={Communication-efficient learning of deep networks from decentralized data},
  author={McMahan, Brendan and Moore, Eider and Ramage, Daniel and Hampson, Seth and y Arcas, Blaise Aguera},
  booktitle={Artificial intelligence and statistics},
  pages={1273--1282},
  year={2017},
  organization={PMLR}
}

@article{li2020federated,
  title={Federated optimization in heterogeneous networks},
  author={Li, Tian and Sahu, Anit Kumar and Zaheer, Manzil and Sanjabi, Maziar and Talwalkar, Ameet and Smith, Virginia},
  journal={Proceedings of Machine learning and systems},
  volume={2},
  pages={429--450},
  year={2020}
}

@article{t2020personalized,
  title={Personalized federated learning with moreau envelopes},
  author={T Dinh, Canh and Tran, Nguyen and Nguyen, Josh},
  journal={Advances in Neural Information Processing Systems},
  volume={33},
  pages={21394--21405},
  year={2020}
}

@inproceedings{thapa2022splitfed,
  title={Splitfed: When federated learning meets split learning},
  author={Thapa, Chandra and Arachchige, Pathum Chamikara Mahawaga and Camtepe, Seyit and Sun, Lichao},
  booktitle={Proceedings of the AAAI Conference on Artificial Intelligence},
  volume={36},
  number={8},
  pages={8485--8493},
  year={2022}
}

@article{he2020group,
  title={Group knowledge transfer: Federated learning of large cnns at the edge},
  author={He, Chaoyang and Annavaram, Murali and Avestimehr, Salman},
  journal={Advances in Neural Information Processing Systems},
  volume={33},
  pages={14068--14080},
  year={2020}
}

@inproceedings{huang2021personalized,
  title={Personalized cross-silo federated learning on non-iid data},
  author={Huang, Yutao and Chu, Lingyang and Zhou, Zirui and Wang, Lanjun and Liu, Jiangchuan and Pei, Jian and Zhang, Yong},
  booktitle={Proceedings of the AAAI conference on artificial intelligence},
  volume={35},
  number={9},
  pages={7865--7873},
  year={2021}
}

@article{diao2020heterofl,
  title={Heterofl: Computation and communication efficient federated learning for heterogeneous clients},
  author={Diao, Enmao and Ding, Jie and Tarokh, Vahid},
  journal={arXiv preprint arXiv:2010.01264},
  year={2020}
}

@article{duan2020fedgroup,
  title={FedGroup: Efficient clustered federated learning via decomposed data-driven measure},
  author={Duan, Moming and Liu, Duo and Ji, Xinyuan and Liu, Renping and Liang, Liang and Chen, Xianzhang and Tan, Yujuan},
  journal={arXiv preprint arXiv:2010.06870},
  year={2020}
}

@article{xu2023parameter,
  title={Parameter-efficient fine-tuning methods for pretrained language models: A critical review and assessment},
  author={Xu, Lingling and Xie, Haoran and Qin, Si-Zhao Joe and Tao, Xiaohui and Wang, Fu Lee},
  journal={arXiv preprint arXiv:2312.12148},
  year={2023}
}

@article{yi2023fedlora,
  title={Fedlora: Model-heterogeneous personalized federated learning with lora tuning},
  author={Yi, Liping and Yu, Han and Wang, Gang and Liu, Xiaoguang},
  journal={arXiv preprint arXiv:2310.13283},
  year={2023}
}

@article{hu2021lora,
  title={Lora: Low-rank adaptation of large language models},
  author={Hu, Edward J and Shen, Yelong and Wallis, Phillip and Allen-Zhu, Zeyuan and Li, Yuanzhi and Wang, Shean and Wang, Lu and Chen, Weizhu},
  journal={arXiv preprint arXiv:2106.09685},
  year={2021}
}

@article{wu2022communication,
  title={Communication-efficient federated learning via knowledge distillation},
  author={Wu, Chuhan and Wu, Fangzhao and Lyu, Lingjuan and Huang, Yongfeng and Xie, Xing},
  journal={Nature communications},
  volume={13},
  number={1},
  pages={2032},
  year={2022},
  publisher={Nature Publishing Group UK London}
}

@inproceedings{collins2021exploiting,
  title={Exploiting shared representations for personalized federated learning},
  author={Collins, Liam and Hassani, Hamed and Mokhtari, Aryan and Shakkottai, Sanjay},
  booktitle={International conference on machine learning},
  pages={2089--2099},
  year={2021},
  organization={PMLR}
}

@article{chen2021bridging,
  title={On bridging generic and personalized federated learning for image classification},
  author={Chen, Hong-You and Chao, Wei-Lun},
  journal={arXiv preprint arXiv:2107.00778},
  year={2021}
}

@String{Computing = "Computing" }

@String{Computer = "{IEEE} Computer" }

@String{Springer = "Springer-Verlag" }

@ArtifactSoftware{R,
    title = {R: A Language and Environment for Statistical Computing},
    author = {{R Core Team}},
    organization = {R Foundation for Statistical Computing},
    address = {Vienna, Austria},
    year = {2019},
    url = {https://www.R-project.org/},
}

\setcounter{table}{0}   
\setcounter{figure}{0}
\setcounter{section}{0}
\setcounter{equation}{0}
\renewcommand{\thetable}{A\arabic{table}}
\renewcommand{\thefigure}{A\arabic{figure}}
\renewcommand{\thesection}{A\arabic{section}}
\renewcommand{\theequation}{A\arabic{equation}}


\section{Conversation Template}
\begin{figure}[!ht]
    \centering
    \includegraphics[scale=0.8]{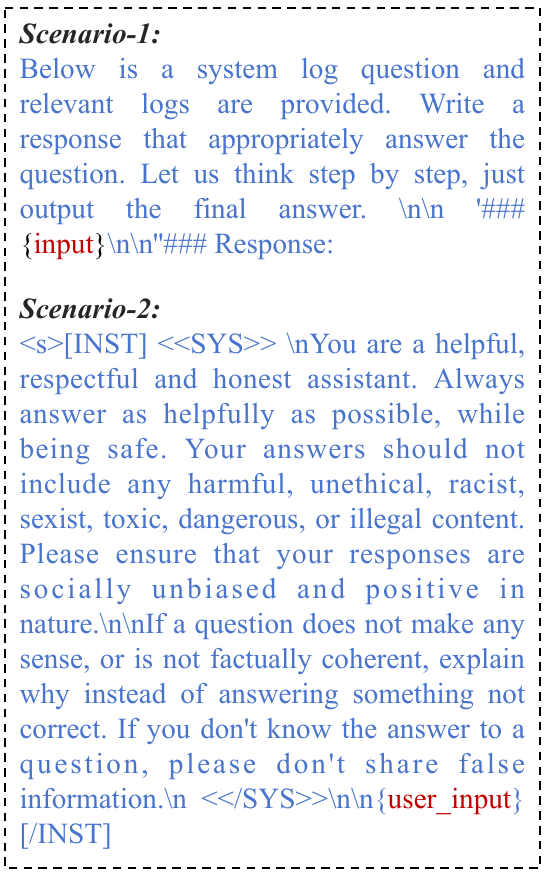}
    \caption{Conversation template with both log-based anomaly detection and medical diagnosis.}
    \label{fig:template}
\end{figure}

As shown in Figure \ref{fig:template}, we show two conversation templates (prompts) integrated with specific input parts (in \textcolor{red}{red}). These templates serve as structured input frameworks for guiding LLMs to generate desired outputs. By providing context and constraints through these prompts, LLMs can effectively process and respond to inputs in a targeted manner, enhancing the quality and relevance of generated outputs. This integration of prompts with input segments facilitates more precise and contextually appropriate responses from LLMs, contributing to improved performance and usability in various natural language processing tasks.
\section{Dataset Example}
\begin{figure}[!ht]
    \centering
    \includegraphics[scale=0.7]{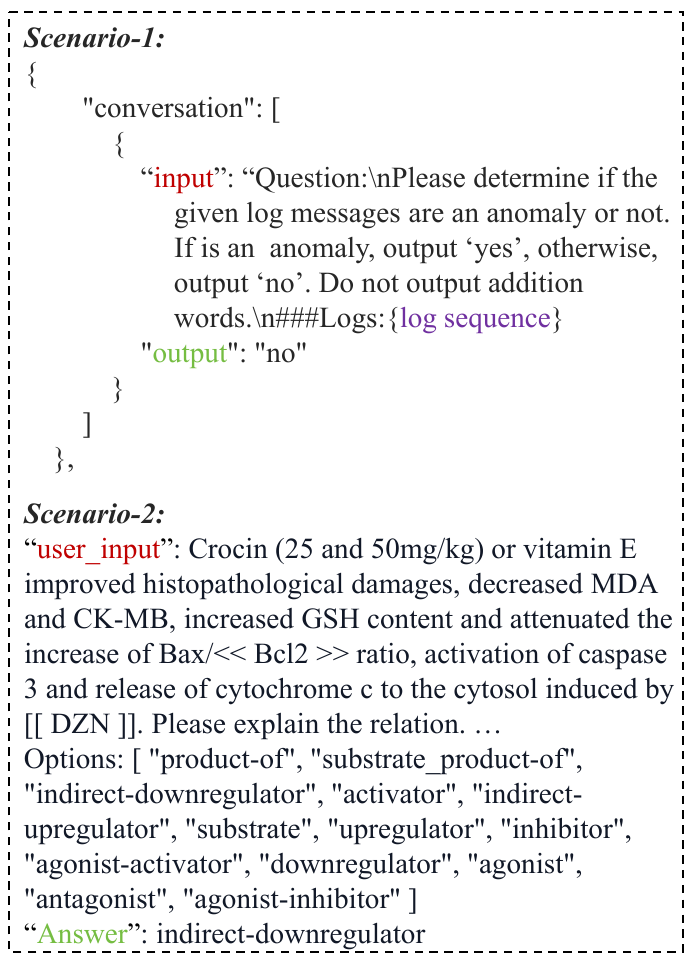}
    \caption{Data example with both log-based anomaly detection and medical diagnosis.}
    \label{fig:data_sample}
\end{figure}

Figure \ref{fig:data_sample} shows the format of the training data in both scenarios. The format is important for the training process. All training data is stored in JSON format, a popular data format. 

\section{Data prepossessing}
\begin{figure}[!ht]
    \centering
    \includegraphics[scale=0.6]{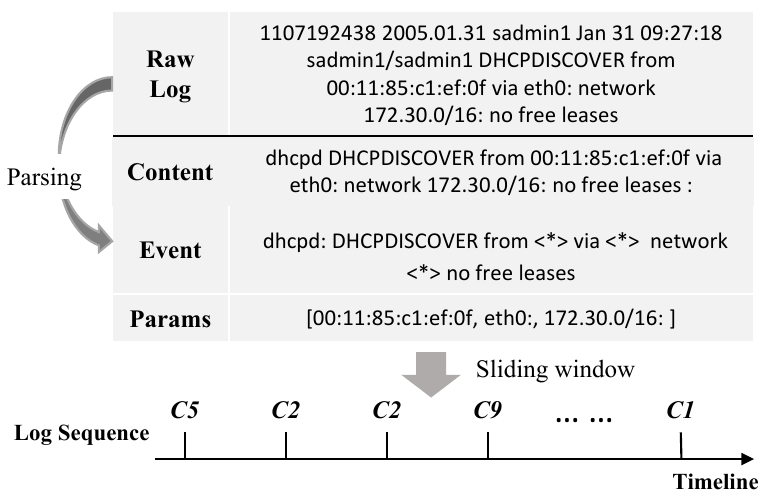}
    \caption{The illustration of log preprocessing.}
    \label{fig:log_process}
\end{figure}

Figure \ref{fig:log_process} is the illustration of preprocessing raw log messages. Log messages typically exhibit a semi-structured format, comprising two main components: a consistent part of the log event, and a variable part containing parameters. To prepare log data for anomaly detection model training, several preprocessing steps are necessary. Initially, both prefix and noise are eliminated through log filtering procedures to obtain the content part. Subsequently, automated log parsing techniques, such as Drain, are applied to extract both log events and parameters. Once parsed, the logs are organized into distinct groups using a fixed-size sliding window (the size default is 50), each comprising multiple log records. These groupings, designated as log sequences, serve as inputs for anomaly detection models and facilitate the identification of various patterns present in the log data. Note that based on existing work, the content part was selected to form the sequence to improve detection performance.

\section{Experiment Environment Details}
As shown in Table \ref{tab:system}, we present the experimental hardware environment and details of LLaMA2-7B. In particular, we used three homogenous devices, so 6 $\times$ V100s were included in total.

\begin{table}[!hb]
\centering
\caption{The evaluated system and LLM specifications. }
\label{tab:system}
\begin{tabular}{ll} 
\toprule
                           & \textbf{System Overview}  \\
                           \midrule
\textbf{CPU}               & 48-core Xeon(R) Gold 6126 CPU @ 2.60GHz       \\
\textbf{GPU}               & 2 $\times$ Tesla V100S       \\
\textbf{Memeory Capacity}  & 187GB DRAM                                    \\
\textbf{Operating System}  & CentOS 7.5.1804                               \\
\textbf{CUDA}              & 12.1                                          \\
\textbf{NVIDIA Driver}     & 530.30.02                                     \\
\textbf{ML framework}      & Pytorch 2.1.2                                 \\
\midrule
                           & \textbf{GPU Specification }                            \\
\midrule
\textbf{CUDA cores}        & 5120                                          \\
\textbf{Memeory Capacity}  & 32GB                                          \\
\textbf{Memeory Bandwidth} & 900GB/s                                       \\
\midrule
                           & \textbf{LLaMA2-7B }                                    \\
\midrule
\textbf{\#Tokens}          & 2.0T                                          \\
\textbf{\#Vocab}           & 32,000                                        \\
\textbf{Seq.Length}        & 4K                                            \\
\textbf{\#Layer}           & 32                                            \\
\bottomrule
\end{tabular}
\end{table}

\end{document}